\documentclass{article}
\usepackage{graphicx}
\usepackage[utf8]{inputenc}
\usepackage{eucal}
\usepackage{amsmath}
\usepackage{hyperref}
\usepackage{epstopdf}

\usepackage{color}
\definecolor{orange}{RGB}{255,127,0}
\definecolor{brown}{RGB}{102,51,0}
\definecolor{myred}{RGB}{192,0,0}
\definecolor{Darkgreen}{RGB}{30,120,30}
\definecolor{Darkblue}{RGB}{0,0,200}
\newcommand{\comment}[1]{}

\newcommand\lsim{\mathrel{\rlap{\lower4pt\hbox{\hskip1pt$\sim$}}
    \raise1pt\hbox{$<$}}}
\newcommand\gsim{\mathrel{\rlap{\lower4pt\hbox{\hskip1pt$\sim$}}
    \raise1pt\hbox{$>$}}}
\newcommand{\ba}{\begin{array}}
\newcommand{\ea}{\end{array}}
\newcommand{\nn}{\nonumber}

\newcommand{\be}{\begin{equation}}
\newcommand{\ee}{\end{equation}}
\newcommand{\bear}{\begin{eqnarray}}
\newcommand{\eear}{\end{eqnarray}}

\newcommand{\ket}{\,\rangle}
\newcommand{\bra}{\langle \,}

\newcommand{\cO}{{\cal O}}

\newcommand{\mL}{\mathcal{L}}

\newcommand{\mB}{\mathcal{B}}

\newcommand{\mF}{\mathcal{F}}

\newcommand{\mO}{\mathcal{O}}
\newcommand{\mP}{\mathcal{P}}

\newcommand{\Frac}[2]{\frac{\displaystyle #1}{\displaystyle #2}}

\title{
Pseudoscalar pole light-by-light contributions to the muon $(g-2)$ in Resonance Chiral Theory}

\author{A. Guevara$^1$, P. Roig$^2$, J. J. Sanz-Cillero$^1$\\
$^1$ {\small Departamento de F\'isica Te\'orica and UPARCOS, Universidad}\\ {\small Complutense de Madrid, Plaza de las Ciencias 1, 28040 Madrid, Spain}\\
$^2$ {\small Centro de Investigaci\'on y de Estudios Avanzados, Apartado}\\{\small
Postal 14-740, 07000, Ciudad de M\'exico, M\'exico}}
\date{}

\begin{document}

\maketitle

\begin{abstract}
    We have studied the $P\to\gamma^\star\gamma^\star$ transition form-factors ($P=\pi^0,\,\eta,\,\eta'$) within a chiral invariant framework that allows us to relate the three form-factors and evaluate the corresponding contributions to the muon anomalous magnetic moment
    $a_\mu=(g_\mu-2)/2$, through pseudoscalar pole contributions.
    We use a chiral invariant Lagrangian to describe the interactions between the pseudo-Goldstones from the spontaneous chiral symmetry breaking and the massive meson resonances. We will consider just the lightest vector and pseudoscalar resonance multiplets. Photon interactions and $U(3)$ flavor breaking effects are accounted for
    in this covariant framework. This article studies the most general corrections of order $m_P^2$ within this setting. Requiring short-distance constraints fixes most of the parameters entering the form-factors,
    consistent with previous determinations.
    The remaining ones are obtained from a fit of these form-factors to experimental measurements in the space-like ($q^2\le0$) region of photon momenta. No time-like  observable is included in our fits.
The combination of data, chiral symmetry relations between form-factors and
high-energy constraints allows us to determine with improved precision the on-shell $P$-pole contribution to the Hadronic Light-by-Light scattering 
of the muon anomalous magnetic moment: we obtain $a_{\mu}^{P,HLbL}=(8.47\pm 0.16)\cdot10^{-10}$ for our best fit.  
This result was obtained excluding BaBar $\pi^0$ data, which our analysis finds in conflict with the remaining experimental inputs.
This study also allows us to determine the parameters describing the $\eta-\eta'$ system in the two-mixing angle scheme and their correlations.  
{Finally, a preliminary rough estimate of the impact of loop corrections ($1/N_C$) and higher vector multiplets (asym) enlarges the uncertainty 
up to $a_\mu^{P,HLbL} = (\, 8.47 \pm 0.16_{\rm sta}  \pm   0.09_{\rm 1/N_C}  {}^{+0.5}_{-0}{}_{\rm asym} ) \cdot 10^{-10}$. 
   }

\end{abstract}

\newpage
\tableofcontents
\newpage

\section{Introduction}

 The electron anomalous magnetic moment, $a_e=(g_e-2)/2$, is the most precisely measured~\cite{PDG} and predicted \cite{Aoyama:2017uqe} observable in nature. There is, however, a greater interest in the muon anomalous magnetic moment, $a_\mu=(g_\mu-2)/2$, since heavy physics beyond the Standard Model (SM) would have an effect of order $(m_\mu/m_e)^2\sim4\times10^4$ times bigger in $a_\mu$ than in $a_e$ \cite{Bijnens:2007pz, Miller:2007kk, Jegerlehner, Lindner:2016bgg}. Similarly, $a_\tau$ should be $\sim280$ times more sensitive to heavy new physics than $a_\mu$. However, the significantly lower mean
lifetime of the $\tau$ lepton makes particularly difficult to measure this property (see e.g. refs.~\cite{Eidelman:2007sb, Eidelman:2016aih}), which by now is still consistent with
zero~\cite{PDG}.\\

The most accurate measurement from Brookhaven \cite{Brookhaven} of $a_\mu$ seems at odds with the SM prediction \cite{Aoyama:2012wk, Gnendiger:2013pva, Davier:2010nc, Hagiwara:2011af, Prades:2009tw, Jegerlehner, Kurz:2014wya, Colangelo:2014qya}
with a 3.5$\sigma$ discrepancy~\cite{PDG} \footnote{The same deviation is quoted in the updated analysis of the leading hadronic vacuum polarization contribution in ref.~\cite{Davier:2017zfy}, while it is 3.7$\sigma$ according to the most recent study in \cite{Keshavarzi:2018mgv}. See ref.~\cite{Actis:2010gg} for a discussion on the Monte Carlo needs for these accurate predictions.} and, both, FNAL muon g-2 and the J-PARC E34 collaborations have announced new experiments that will reduce the current Brookhaven error by, at least, a factor 4 \cite{FNAL,JPARC}. Therefore, it becomes necessary to make a more accurate prediction for this observable with reduced uncertainty so as to be able to confront
the forthcoming results from both collaborations with a SM result of comparable accuracy.\\

The main source of uncertainty comes from the hadronic contributions, so these are where our activity should focus on (see in ref.~\cite{Melnikov:2001uw} an
early discussion of this issue). All such contributions can be divided into two, the Hadronic Vacuum Polarization (HVP) and the Hadronic Light-by-Light Scattering
(HLbL). The HVP can be completely data-driven through dispersion relations \cite{Brodsky:1967sr, Gourdin:1969dm}~\footnote{Also lattice QCD simulations have
recently managed to give accurate determinations. See, e.g., refs.~\cite{latticeHVP}.}, while the latter cannot be obtained completely in this way yet.
However, there have been remarkable advances in determining the HLbL part in a model independent way, by means of Lattice QCD \cite{Blum:2016lnc,Blum:2017cer,Green:2015sra} 
and dispersion relations \cite{Bern,Mainz}. 
Therefore, it is the HLbL contribution to the muon anomalous magnetic
moment, $a_\mu^{HLbL}$, which calls for a dedicated theory effort \cite{Blum:2013xva}.\\

We will focus on the leading contribution to the HLbL
\cite{Jegerlehner}, which is given by the pseudoscalar exchange \footnote{This dominance is not fully understood from a first-principles derivation \cite{deRafael:1993za, Knecht:2001qg}.}, $a_\mu^{P,HLbL}$. 
To evaluate such contribution, a necessary ingredient is the $P\to\gamma^\star\gamma^\star$ transition form-factor (TFF), $\mathcal{F}_{P\gamma^\star\gamma^\star}$, which cannot be computed analytically in the underlying theory of strong interactions. 
{However, as for the HLbL part, there have also been remarkable advances in determining the TFF in a model independent way by means of Lattice QCD \cite{Gerardin:2016cqj} 
and dispersion relations \cite{DispersionRel}.}
We calculate it by means of the extension of Chiral Perturbation Theory ($\chi$PT)~\cite{ChPT} that incorporates the lightest resonance multiplets in a chiral invariant framework~\cite{RCT}, called Resonance Chiral Theory (R$\chi$T). We work within the large--$N_C$ limit and assume a $1/N_C$ expansion, in such a way that we have a spontaneous chiral symmetry breaking pattern $U(3)_L\times U(3)_R/U(3)_{L+R}$, which gives place to a nonet of chiral (pseudo) Goldstones. We consider a R$\chi$T Lagrangian, $\mL_{R\chi T}$, that includes $U(3)$ symmetric operators and terms that introduce quark mass corrections in the even and odd-intrinsic parity sectors~\cite{Cirigliano:2006hb,MassSplitting,VVP,Kampf}. Previous works~\cite{Kampf} and~\cite{us} analyzed $U(3)$ symmetric TFFs, given in the chiral limit. The novelty of the present approach is that the TFFs are studied beyond the massless pseudo-Goldstone limit, accounting for its leading order corrections in powers of the pseudo-Goldstone bosons squared masses, $m_P^2$, which explicitly break $U(3)$ flavor symmetry. We will see that all but eight parameters (including among them the four $\eta-\eta'$ mixing parameters) are fixed by short distance constraints. These eight unknown couplings will be determined through a fit to the experimental data in the space-like region of photon four-momenta $q^2\le0$ (note that the pseudoscalar pole contribution to $a_\mu$ can be written in terms of just the space-like TFFs~\cite{Knecht}). Since the data for the $\pi^0$ transition form-factor given by BaBar collaboration \cite{BaBar} seems to be at odds with its Brodsky-Lepage high-energy limit~\cite{BL} and the $\eta$ and $\eta'$ TFFs, we will discuss its validity and consistency with other data. We will perform various alternative analyses and take as our reference fit the one without this set of data.\\

Our approach intends to introduce the $U(3)$ breaking through a chiral invariant Lagrangian, where quark mass effects enter in a covariant way. This gives slightly simpler expressions for these form-factors and the short-distance constraints for the Lagrangian parameters, contrary to the strategy followed in ref.~\cite{Czyz}, which allows a completely general $U(3)$ breaking pattern. Despite giving an accurate description of data, is a little less straightforward to employ and does not rely on chiral symmetry (where quark mass correction must enter through a more restricted pattern). Likewise, ref.~\cite{Czyz} also used data from both time- and space-like regions, whereas the present article will only rely on space-like data. We will work in the large--$N_C$ limit and hadron loop effects will be neglected, so time-like observables (e.g., partial widths) will be excluded from our fits, as they require a dedicated analysis of the $1/N_C$ corrections. In addition, photon radiative corrections to $P\to\gamma^{(\star)}\gamma^{(\star)}$ decays also play a non negligible role~\cite{NLODD}.\\

The paper is organized as follows: In section \ref{Lagrangian}, we recall the relevant pieces of the Resonance Chiral Lagrangian needed for our study. Particularly, we explain the flavor-breaking corrections that we include for the first time in this kind of analysis (see, however, ref.~\cite{Ametller:1991jv} for a related discussion within vector-meson dominance, constituent-quark loops, the QCD-inspired interpolation by Brodsky-Lepage, and Chiral Perturbation Theory). Then, in section \ref{Computations} we collect our results for the transition form-factors including the contribution from intermediate vector and pseudoscalar resonances. In section \ref{SD}
we discuss the short-distance constraints on the Resonance Chiral Lagrangian parameters that are obtained by demanding the QCD short-distance behaviour to the VVP Green's function and the pseudo-Goldstone form-factors ($P$--TFF). Our fits to data are presented in section \ref{Fits}, where we also quote our results for the parameters describing the $\eta-\eta'$ mixing in the double angle scheme, including their correlations.
We also compare our approach to other recent articles on the subject in section \ref{Comparisons}.
Finally, the branching ratio predictions for $P\to\gamma^{(\star)}\gamma^{(\star)}$ processes are compared to their respective experimental measurements in section \ref{Predictions}.  The corresponding $P$ pole contributions to $a_\mu$ are obtained in section \ref{a_mu}, with a careful
statistical treatment and highlighting the influence of $\pi^0$ transition form-factor BaBar data. Finally, our conclusions are summarized in section \ref{Concl}. The three appendices collect, respectively, the Wess-Zumino--Witten Lagrangian, the relevant formulae used to evaluate the $P$ pole contribution to $a_\mu$, and the correlation matrices for the two alternative fits considered in our analysis.

\section{R$\chi$T Lagrangian}
\label{Lagrangian}

\subsection{Relevant operators for the TFF}

   {
   In modeling the TFF we make use of R$\chi$T, 
   an extension of $\chi$PT that also includes the lightest resonance multiplets~\cite{RCT}.
   The Wess-Zumino-Witten action \cite{WZW} describes the local $P\gamma\gamma$ interaction vertex, whereas a dressed photon description is
   needed to assure the ultraviolet convergence of the pseudo-Goldstone exchange contribution to $a_\mu^{HLbL}$~\cite{Jegerlehner}. Within our
   approach, this is done by considering the vector resonances exchange between the pseudo-Goldstone and the final state photons. To include such
   interactions one needs to consider the odd-intrinsic parity sector with operators including a pseudo-Goldstone and, either two vector resonances
   or one vector resonance and one external photon. \\

   The complete odd-intrinsic parity basis involving pseudo-Goldstones and either two vector resonances or a vector resonance 
   and a photon was given in~\cite{Kampf}. We will rely, however, on the operators given in ref. \cite{VVP}, 
   since these conform a complete basis for describing vertices
   involving only one pseudo-Goldstone. Also, this basis is simpler for the problem at hand 
   (optimized for the study of this type of processes) and, as proven in~\cite{SDConst}, 
   equivalent to the complete basis of Ref.~\cite{Kampf} for a single pseudo-Goldstone field.
   In order to compare and employ results from~\cite{Kampf}, pseudoscalar resonances $P'$ are also included in
   our description of the TFF. Ref.~\cite{us} found that the effect of the latter on the form-factors was equivalent to adding a second vector multiplet.  
   Such $P'$ interactions are taken from ref.~\cite{Kampf} within the chiral limit, 
   which give their first corrections to the TFF at $\cO(m_P^2)$ and will be addressed in
   subsection~\ref{sec:U3-breaking}.\\

   }

The Lagrangian describing the interaction between the lightest multiplet of pseudoscalars (the chiral pseudo-Goldstones) and massive meson resonances, $R$, can be organized
according to the number of heavy fields:
\bear
\mL&=& \mL_{\rm non-R} +\sum_R \left(\mL_R^{\rm Kin}+\mL_R \right)+  \sum_{R,R'} \mL_{RR'}  +\sum_{R,R',R''} \mL_{RR'R''} + ...
\eear
We will not consider operators with four or more resonance fields since
they cannot contribute to the $P\to\gamma^\star\gamma^\star$ TFFs at the tree-level --considered in this analysis--.
We will describe the spin--1 resonance fields in the antisymmetric tensor representation~\cite{RCT}. In this formalism, as a general feature, the simplest operators in the even-intrinsic parity sector contain an $\cO(p^2)$ chiral tensor (with two derivatives or equivalent scales)~\cite{RCT} in addition to the resonance field,
while in the odd-intrinsic parity sector the light-pseudoscalar tensor accompanying the resonance fields usually starts at $\cO(p^4)$~\cite{VVP,Kampf}.
Furthermore, the R$\chi$T Lagrangian will be divided in even and odd-intrinsic parity sectors, the latter containing the anomalous Wess-Zumino-Witten Lagrangian, $\mL_{WZW}$. We will present now the R$\chi$T operators that provide the relevant interactions between photons, $\gamma$, pseudo-Goldstones, $\phi^a$, and vector resonances, $V$:
\begin{itemize}
\item{\bf Operators without resonance fields:}\\
The operators in  $\mL_{\rm non-R}$ only contain pseudo-Goldstone fields $\phi^a$
and start in the even-intrinsic parity sector at $\cO(p^2)$ (given by Gasser and Leutwyler's $\mL^{\cO(p^2)}_{\chi PT}$ Lagrangian) and at $\cO(p^4)$ for the anomalous odd-intrinsic parity sector (where the lowest order contribution is provided by $\mL_{WZW}$).
However, as we want to study the most general possible $U(3)$ breaking we will also consider operators with the structure of the odd-intrinsic parity $\cO(p^6)$ $\chi$PT Lagrangian~\cite{Bijnens}. Thus our non-resonant part of the R$\chi$T Lagrangian is given by
\bear
\mL_{\rm non-R}^{\rm even}&=&\frac{F^2}{4}\langle u_\mu u^\mu+\chi_+\rangle\, ,
\nn\\
\mL_{\rm non-R}^{\rm odd}&=&\mL_{WZW} + \sum_{j=7,8,22}C_j^W \mO_j^W\, ,
\label{eq:L-nonR}
\eear
where $\langle A\rangle$ stands for the trace in flavour space of $A$, the operators $\mO_j^W$ can be found in table~\ref{tab:OjW} and the well-known --though lengthy-- form of $\mL_{WZW}$ is given in App.~\ref{app:WZW}~\cite{WZW,Bijnens}. Although this non-resonant Lagrangian was considered in the chiral and large--$N_C$ limit VVP Green's function analysis~\cite{Kampf}, an appropriate description of physical $P\to\gamma^\star\gamma^\star$ processes requires further pseudo-Goldstone bilinear terms not shown above~\cite{Kaiser:2000gs,Kaiser:2000ck,Guo:2015xva}, which dress the $\phi^a$ wave-functions and induces the $\eta-\eta'$ mixing. This details are discussed in the later section~\ref{sec:U3-breaking}.
In order to explore the $U(3)$ breaking in the $\eta-\eta'$ system in more generality we have allowed here the presence of the double-trace operator $C_8^W$. Although it is subleading in $1/N_C$ and it should be dropped from our analysis, $1/N_C$ corrections play an important role in the description of the $\eta-\eta'$ mixing so one could argue that this operator might be numerically relevant.
We will show, nonetheless, that after demanding that the TFFs follow the high-energy QCD behaviour this subleading coupling $C_8^W$ vanishes.
We want to emphasize that $\mL_{\rm non-R}$ in eq.~(\ref{eq:L-nonR}) is not the low-energy $\chi$PT Lagrangian; it belongs to R$\chi$T Lagrangian which describes the meson interactions in the range of high and intermediate energies and the $C_j^W$ and $C_j^{W,\, {\rm \chi PT}}$ couplings must not be confused.

\begin{table}[!t]
    \centering
    \begin{tabular}{c c }\hline\hline   \\\hline
    $\mO_7^W$ &  $i \epsilon_{\mu\nu\alpha\beta} \bra \chi_- f_+^{\mu\nu} f_+^{\alpha\beta}\ket$
    \\
    $\mO_8^W$& $ i \epsilon_{\mu\nu\alpha\beta} \bra \chi_-\ket \, \bra  f_+^{\mu\nu} f_+^{\alpha\beta}\ket$
    \\
    $\mO_{22}^W$& $i \epsilon_{\mu\nu\alpha\beta} \bra u^\mu \{ \nabla_\rho f_+^{\rho\nu}, f_+^{\alpha\beta}\} \ket$
    \\\hline\hline
    \end{tabular}
    \caption{{\small Relevant non-resonant odd-intrinsic parity operators.}}
    \label{tab:OjW}
\end{table}

\item{\bf Operators with one vector resonance field:}\\
At tree-level, the $P$--TFF will have contributions from $V-\gamma$ and $V-\gamma-\phi^a$ vertices,
with the vector multiplet $V_{\mu\nu}=\sum_{a=0}^8\frac{1}{\sqrt{2}} \lambda^a V_{\mu\nu}^a$ described in the antisymmetric tensor formalism~\cite{RCT}~\footnote{The operator $\Delta \mL=\frac{i}{2\sqrt{2}} G_V\bra V_{\mu\nu} [u^\mu,u^\nu]\ket$~\cite{RCT} is not included here in $\mL_V^{\rm even}$ since it does not contribute to the studied processes at tree-level. The same applies to other Lagrangian operators from refs.~\cite{RCT,Kampf,Cirigliano:2006hb} not quoted here.}:
\bear
\mL_R^{\rm Kin}&=& -\Frac{1}{2} \bra \nabla_\lambda V^{\lambda \nu} \nabla^\rho V_{\rho \nu} \ket
+\Frac{1}{4}M_V^2 \bra V_{\mu\nu} V^{\mu\nu}\ket \, ,
\nn\\
\mL_V^{\rm even} &=& \Frac{F_V}{2\sqrt{2}}\bra V_{\mu\nu} f_+^{\mu\nu}\ket + \Frac{\lambda_V}{\sqrt{2}}\bra V_{\mu\nu} \{  f_+^{\mu\nu}, \chi_+\}\ket \, ,
\nn\\
\mL_V^{\rm odd} &=& \sum_{j=1,2,3,5,6} \Frac{c_j}{M_V}\mO^j_{VJP}\, .
\eear
A quark mass correction to the  $V-\gamma$ transitions is allowed in $\mL_V^{\rm even}$ to account for the $U(3)$ breaking in this vertex,
where the only single-trace operator at $\cO(m_P^2)$ is given by the $\lambda_V$ term, given in ref.~\cite{Cirigliano:2006hb} by $\mO_6^V$ with coupling $\lambda_6^V=\lambda_V/\sqrt{2}$.
Table~\ref{tab:VJP} provides the full list of $\mO^j_{VJP}$ operators~\cite{VVP},
containing those ($j=1,2,3,5,6$) that contribute to the $V-\gamma-\phi^a$ vertex. Notice that the $c_3$ term is the only one that explicitly breaks the $U(3)$ symmetry, via the tensor $\chi_-$, proportional to the quark masses. Nonetheless, the other $c_j$ couplings will also enter in the $U(3)$ breaking contributions to the TFFs once the external pseudo-Goldstones are set on-shell and their wave function renormalizations and mixings are taken into account.

\begin{table}[!t]
    \centering
    \begin{tabular}{c c }\hline\hline   \\\hline
   $ \mathcal{O}^1_{VJP}$&$\varepsilon_{\mu\nu\rho\sigma}\langle\left\{V^{\mu\nu},f^{\rho\alpha}_+\right\}\nabla_\alpha u^\sigma\rangle$\\
  $\mathcal{O}^2_{VJP}$&$\varepsilon_{\mu\nu\rho\sigma}\langle\left\{V^{\mu\alpha},f^{\rho\sigma}_+\right\}\nabla_\alpha u^\nu\rangle$\\
  $\mathcal{O}^3_{VJP}$&$i\varepsilon_{\mu\nu\rho\sigma}\langle\left\{V^{\mu\nu},f^{\rho\sigma}_+\right\}\chi_-\rangle$\\
  $\mathcal{O}^4_{VJP}$&$i\varepsilon_{\mu\nu\rho\sigma}\langle V^{\mu\nu}\left[f^{\rho\sigma}_-,\chi_+\right]\rangle$\\
  $\mathcal{O}^5_{VJP}$&$\varepsilon_{\mu\nu\rho\sigma}\langle\left\{\nabla_\alpha V^{\mu\nu},f^{\rho\alpha}_+\right\}u^\sigma\rangle$\\
  $\mathcal{O}^6_{VJP}$&$\varepsilon_{\mu\nu\rho\sigma}\langle\left\{\nabla_\alpha V^{\mu\alpha},f^{\rho\sigma}_+\right\}u^\nu\rangle$\\
  $\mathcal{O}^7_{VJP}$&$\varepsilon_{\mu\nu\rho\sigma}\langle\left\{\nabla^\sigma V^{\mu\nu},f^{\rho\alpha}_+\right\}u_\alpha\rangle$    \\\hline\hline
    \end{tabular}
    \caption{{\small Full list of $\mO^j_{VJP}$ operators~\cite{VVP},
containing those ($j=1,2,3,5,6$) that contribute to the $V-\gamma-\phi^a$ vertex.}}
    \label{tab:VJP}
\end{table}

\item{\bf Operators with two vector resonance fields: }\\
Operators with two vector fields give quark mass corrections to the resonance mass term (although we have preferred to keep it as a perturbation instead of including it in $\mL_V^{\rm Kin}$)
and contribute to the $V-V-\phi^a$ vertices:
\bear
\mL_{VV}^{\rm even} &=& - e_m^V \bra V_{\mu\nu}V^{\mu\nu}\chi_+\ket \, ,
\nn\\
\mL_{VV}^{\rm odd} &=& \sum_{j=1,2,3} d_j \mO_{VVP}^j \, .
\eear
The quark mass splitting induces a $U(3)$ mass breaking in the vector nonet, in fair agreement with the phenomenology~\cite{MassSplitting}. We note that in ref.~\cite{MassSplitting} the $e_m^V$ Lagrangian is loosely written as $e_m^R\bra RR\chi_+\ket$ for a generic resonance.
Table~\ref{tab:VVP} provides the full list of $\mO^j_{VVP}$ operators~\cite{VVP},
containing those ($j=1,2,3$) that contribute to the $V-V-\phi^a$ vertex. Only the $d_2$ operator breaks explicitly $U(3)$.

\begin{table}[!t]
    \centering
    \begin{tabular}{c c }\hline\hline   \\\hline
    $\mathcal{O}^1_{VVP}$&$\varepsilon_{\mu\nu\rho\sigma}\langle\left\{V^{\mu\nu},V^{\rho\alpha}\right\}\nabla_\alpha u^\sigma\rangle$\\
  $\mathcal{O}^2_{VVP}$&$i\varepsilon_{\mu\nu\rho\sigma}\langle\left\{V^{\mu\nu},V^{\rho\sigma}\right\}\chi_-\rangle$\\
  $\mathcal{O}^3_{VVP}$&$\varepsilon_{\mu\nu\rho\sigma}\langle\left\{\nabla_\alpha V^{\mu\nu},V^{\rho\alpha}\right\}u^\sigma\rangle$\\
  $\mathcal{O}^4_{VVP}$&$\varepsilon_{\mu\nu\rho\sigma}\langle\left\{\nabla^\sigma V^{\mu\nu},V^{\rho\alpha}\right\}u_\alpha\rangle$
  \\\hline\hline
    \end{tabular}
    \caption{{\small Full list of $\mO^j_{VVP}$ operators~\cite{VVP},
containing those ($j=1,2,3$) that contribute to the $V-V-\phi^a$ vertex.  }}
    \label{tab:VVP}
\end{table}

\end{itemize}

The chiral building blocks are defined as~\cite{Bijnens}
\begin{eqnarray}
    u_\mu&=&i\left[u^\dagger(\partial_\mu-ir_\mu)u - u(\partial_\mu-i\ell_\mu)u^\dagger\right]\, ,
    \nonumber\\
    \Gamma_\mu&=& \frac{1}{2}\left[u^\dagger(\partial_\mu-ir_\mu)u + u(\partial_\mu-i\ell_\mu)u^\dagger \right]\, ,
    \nn\\
    \nabla_\mu\cdot &=& \partial_\mu \cdot + [\Gamma_\mu, \cdot]\, ,
    \qquad
    f_\pm^{\mu\nu} = u F_L^{\mu\nu} u^\dagger \pm u^\dagger F_R^{\mu\nu} u\, ,
    \nn\\
    \chi_{\pm}&=&u^\dagger\chi u^\dagger\pm u\chi^\dagger u, \hspace*{5ex}\chi=2B_0(s+ip) \, ,
\nn\\
u&=& \exp{\left(\Frac{i \Phi}{\sqrt{2}F}\right)}\, ,
\qquad \Phi=\sum_{a=0}^8\Frac{ \lambda_a \phi^a}{\sqrt{2}}\,,
\end{eqnarray}
where the $\phi^a$ fields provide the $U(3)$ nonet of the chiral pseudo-Goldstone fields (the $\eta'$ becomes a chiral Goldstone under the chiral and large--$N_C$ limits), the scalar-pseudoscalar densities are set to the quark masses, $s+ip=$diag$(m_u,m_d,m_s)$ (isospin symmetry will be assumed from now on, so $m_u=m_d\equiv m_q\neq m_s$), and $F$ is the pion decay constant ($F_\pi=92.2$~MeV~\cite{PDG}) in the chiral limit. The left and right sources are respectively
$\ell_\mu=v_\mu-a_\mu=e Q A_\mu+...$ and $r_\mu=v_\mu+a_\mu=e Q A_\mu+...$ (with $A_\mu$ the photon field, $Q=$diag$\left(\frac{2}{3},-\frac{1}{3},-\frac{1}{3}\right)$), $F_{L}^{\mu\nu}=\partial^\mu\ell^\nu-\partial^\nu\ell^\mu - i[\ell^\mu,\ell^\nu]= e Q F^{\mu\nu}+...$ and $F_R^{\mu\nu}=F_L^{\mu\nu}|_{\ell\to r}= e Q F^{\mu\nu}+...$ are their associated field strengths (with the photon field strength tensor $F^{\mu\nu}=\partial^\mu A^\nu -\partial^\nu A^\mu$),
and the dots stand for gauge boson fields not contributing to the $P\to\gamma^\star\gamma^\star$ TFF.
For the Levi-Civita tensor we use the standard convention $\epsilon_{0123}=1$.

The main contribution to the TFF is provided by the previous operators, as the pseudoscalar resonances $P'$ can only enter via the $P'-\phi^a$ mixing, suppressed by the ratio $m_P^2/M_{P'}^2$ of the light pseudo-Goldstone and heavy pseudoscalar resonance masses.
However, the $P'$ multiplet is crucial to recover the right behaviour prescribed by the OPE for the VVP Green's function~\cite{Kampf,SDConst}. In order to use the short-distance constraints therein, for consistency, one needs to also include the pseudoscalar resonance operators in~\cite{Kampf}
that contribute to the TFF~\cite{RCT,Kampf}:
\begin{eqnarray}
\Delta \mL_{P'}^{\rm even} &=&\Frac{1}{2} \bra \nabla_\mu P'\, \nabla^\mu P'\ket + id_m \bra P'\chi_-\ket \, ,
\\
\Delta \mL_{P'}^{\rm odd} &=&\varepsilon_{\mu\nu\alpha\beta}\langle \kappa_5^P\{f_+^{\mu\nu}, f_+^{\alpha\beta}\}P'+\kappa_3^{PV}\{V^{\mu\nu},f_+^{\alpha\beta}\}P'+\kappa^{PVV}V^{\mu\nu}V^{\alpha\beta}P' \rangle\,,
\nn
\end{eqnarray}
with $P'=\sum_{a=0}^8 \lambda_a P^{'\, a}/\sqrt{2}$.
Our final results will crucially rely on the asymptotic behaviour imposed
by QCD on the TFFs at high-energies~\cite{BL}.

\subsection{$U(3)$ breaking in the $\Phi$ and $V$ nonets}
\label{sec:U3-breaking}

   A more accurate analysis of the $P$--TFF can be done by including corrections up to $\cO(m_P^2)$,
   with $m_P$ referring to the mass of the pseudo-Goldstones. In order to do this, one needs to take into account all possible corrections to interaction vertices, resonance masses and field renormalizations.

   For the vector resonance nonet we will assume an ideal mixing, such that
   $(V^{\mu\nu}_{11},V^{\mu\nu}_{22},V^{\mu\nu}_{33}) = \left( (\rho^{0\, \mu\nu}+\omega^{\mu\nu})/\sqrt{2} ,  (-\rho^{0\, \mu\nu}+\omega^{\mu\nu})/\sqrt{2} , \phi^{\mu\nu}\right) $,
   as prescribed by the large--$N_C$ limit.
   We will include only the lowest order terms in $m_{q/s}$ that break the 
   $U(3)$ symmetry.
   The quark mass corrections to the vector masses stemming from the operator of $\mL^\text{even}_{VV}$ do not modify this mixing.
   In the large--$N_C$ and isospin limits, this Lagrangian yields the vector resonance mass pattern~\cite{MassSplitting},
 \begin{subequations}
  \begin{align}
   M_\rho^2=&M_V^2 - 4e^V_mm_\pi^2\, ,
   \\
   M_\omega^2=&M_V^2 - 4e^V_mm_\pi^2\, ,
   \\
   M_{K^*}^2=&M_V^2 - 4e^V_mm_K^2\, ,
   \\
   M_\phi^2=&M_V^2 - 4e^V_m  \Delta^2_{2 K\pi} \, ,
  \end{align}\label{Vmasses}
 \end{subequations}
 in fair agreement with the phenomenology,
 with $\Delta^2_{2 K\pi}=2 m_K^2-m_\pi^2$. The $K^*$ mass is provided to complete the information on the $U(3)$ splitting of the vector nonet, although it will not be relevant for this work, obviously.
 Similar relations can be obtained for the pseudoscalar resonances (see ref.~\cite{MassSplitting}), however these will not be needed. These effects would only correct the amplitudes at $\cO(m_P^4)$
 and are beyond the scope of this work.

In the same way, the $\lambda_V$ term in $\mL_{V}^\text{even}$ Lagrangian will introduce an $U(3)$ breaking in the $V-\gamma$ transitions, such that we will have to make the replacements:
 \begin{subequations}
  \begin{align}
\rho-\gamma:\qquad    &F_V\quad \longrightarrow\quad F_V + 8 m_\pi^2 \lambda_V,\\
\omega-\gamma:\qquad    &F_V\quad \longrightarrow\quad F_V + 8 m_\pi^2 \lambda_V,\\
\phi-\gamma:\qquad    &F_V\quad \longrightarrow\quad F_V + 8 \Delta_{2K\pi}^2 \lambda_V\, .
  \end{align}\label{FVsplitting}
 \end{subequations}

  To compute the transition $P\to\gamma\gamma$ one has to establish the pseudo-Goldstone mass eigenstates and, in particular, the $\eta-\eta'$ mixing. For this we will use a
 two-angle mixing scheme \cite{Kaiser:2000gs,eta-etap} consistent with the large $N_C$ limit of QCD \cite{largeNC}, where the octet and singlet bare fields $\phi^a$ are related to the physical fields through
 \begin{equation}
  (\Phi_{11},\Phi_{22},\Phi_{33})
  =\left(\frac{C_\pi\pi^0+C_q\eta+C_q'\eta'}{\sqrt{2}},\frac{-C_\pi\pi^0+C_q\eta+C_q'\eta'}{\sqrt{2}},-C_s\eta+C_s'\eta'\right),
 \end{equation}
 where $\Phi=\sum_{a=0}^8\frac{1}{\sqrt{2}}\phi^a\lambda^a$ is the pseudo-Goldstone matrix
 in terms of the bare pseudoscalar fields $\phi^a$.
 These fields are related to the physical ones through a rescaling of the form
 $\phi^3=(\Phi_{11}-\Phi_{22})/\sqrt{2}=C_\pi \pi^0$, with $C_\pi= F/F_\pi$ in the large--$N_C$ limit~\cite{Bernard:1991zc,SanzCillero:2004sk,Guo:2014yva}. In the most general case, when loops and other corrections are taken into account $C_\pi^2$ is just the wave-function renormalization $Z_\pi$.

 Within the two-angle mixing scheme, the mixing constants are parametrized in the most general form through
 \begin{subequations}
  \begin{align}
   C_q:=&\frac{F}{\sqrt{3}\cos(\theta_8-\theta_0)}\left(\frac{\cos\theta_0}{f_8}-\frac{\sqrt{2}\sin\theta_8}{f_0}\right),\\
   C_q':=&\frac{F}{\sqrt{3}\cos(\theta_8-\theta_0)}\left(\frac{\sqrt{2}\cos\theta_8}{f_0}+\frac{\sin\theta_0}{f_8}\right),\\
   C_s:=&\frac{F}{\sqrt{3}\cos(\theta_8-\theta_0)}\left(\frac{\sqrt{2}\cos\theta_0}{f_8}+\frac{\sin\theta_8}{f_0}\right),\\
   C_s':=&\frac{F}{\sqrt{3}\cos(\theta_8-\theta_0)}\left(\frac{\cos\theta_8}{f_0}-\frac{\sqrt{2}\sin\theta_0}{f_8}\right)\, ,
  \end{align}
 \end{subequations}
 in terms of two couplings $f_{8/0}$ and two mixing angles $\theta_{8/0}$~\cite{mixing}. The physical fields can be identified through the mixing parameters as a linear combination of the `bare' $SU(3)$ octet (bare $\eta_8=\phi^8$) and singlet (bare $\eta_0=\phi^0$) fields in $\Phi$. In the large $N_C$ limit, one has $\eta_q=\eta$ and $\eta_s=\eta'$, where the former is an isosinglet combination $u\bar{u}+d\bar{d}$ and the latter is a pure $s\bar{s}$ state.
 These parameters have been computed in $U(3)$ large--$N_C$ $\chi$PT at LO~\cite{Kaiser:2000gs}
 (where $\theta_8=\theta_0\approx  -20^\circ$ and $f_8=f_0=F$), next-to-leading order (NLO)~\cite{Kaiser:2000gs} and next-to-next-to-leading order (NNLO)~\cite{Guo:2015xva}. These calculations point out an important issue in the treatment of the $\eta$ and $\eta'$ mesons with the large--$N_C$ and/or chiral limit:
 $1/N_C$ and quark mass corrections to the $U(3)$ pseudo-Goldstone properties happen to be of a similar order of magnitude and compete with each other. For instance, the lowest contribution to the $\eta'$ mass is dominated by the $\cO(N_C^{-1} m_q^0)$ contribution from the $U(1)_A$~chiral anomaly whereas the $\cO(N_C^0 m_q^1)$ contributions mostly determine the $\pi^0$ mass.
 It is illustrative to observe the two limits separately: in the large--$N_C$ limit,
 the physical states are given by the ideal mixing angles $\theta_{8/0}=-\arcsin{\left(\sqrt{2/3}\right)} \approx - 55^\circ$
 and couplings $f_{8/0}=F$,
 which imply $C_q=C_s'=1$ and $C_q'=C_s=0$, all up to quark mass corrections;
 in the chiral limit, $\theta_0=\theta_8=0$, $f_8=F$ and $f_0=F\, [1+\cO(N_C^{-1})]$,
 which imply $C_q=\sqrt{1/3}$, $C_q'=\frac{F}{f_0}\sqrt{2/3} $,  $C_s=\sqrt{2/3}$ and $C_s'=\frac{F}{f_0}\sqrt{1/3} $.\\

 Therefore, in the present work, we will always keep the full mixing coefficients $C_{q/s}^{(')}$, not expanded; on the other hand,
 the large--$N_C$ limit will be assumed everywhere else in our space-like TFF analysis and resonance widths, meson loops or multi-trace
 R$\chi$T operators will be considered negligible for our study. Obviously, these subleading effects must be properly taken into account
 in the analysis of time-like observables (vector and pseudoscalar branching ratios, $e^+e^-\to P\gamma$ production, etc.). We are, however,
 modelling the large--$N_C$ limit of $R\chi T$ with the infinite tower of resonant states truncated at the first multiplet. Still, we  expect
 that this approximation will affect very mildly our prediction for $a_\mu^{P,HLbL}$. The reason for this is that the integration kernel of
 the these contributions is completely dominated by the $[0.1,1]$ GeV$^2$ region for both photon virtualities~$q_{1,2}^2$~\cite{Knecht},
 which suppresses the contributions from higher resonance excitations. {Nonetheless we will use the $[0,100]$ GeV$^2$ region for both
 photon momenta in the form-factor when computing $a_\mu^{P,HLbL}$. According to our results, $\sim85\%$ of the whole contribution comes
 from the $[0.1,1]$ GeV$^2$ region for both photon virtualities. Furthermore, if the upper limit in both squared momenta is set to 400 GeV$^2$,
 the prediction for $a_\mu^{P,HLbL}$ changes by less than 0.005\% with respect to our reference value.}
 In addition to this, we will see in section \ref{Fits} that current
 TFF data (extending up to roughly $35$ GeV$^2$) do not show any hint of a sizable contribution from excited resonances and that their effect
 can be captured by a small shift of the lightest vector resonance parameters. It is illustrative to compare the present results with more
 involved studies including higher resonance poles~\cite{Czyz,Knecht,Masjuan:2017tvw}.

\section{Transition form-factors in R$\chi$T}\label{Computations}

The transition amplitude $P(p)\to\gamma^\star(q_1,\epsilon_1)\gamma^\star(q_2,\epsilon_2)$ with an on-shell light pseudoscalar and two photons with virtualities $q_1^2$ and $q_2^2$ (polarizations $\epsilon_1^\star(q_1)$ and $\epsilon_2^\star(q_2)$) can be parametrized in terms of a scalar function, the form-factor $\mF_{P\gamma^\star\gamma^\star}(q_1^2,q_2^2)$, in the form
\begin{eqnarray}
  \mathcal{M}_{P\gamma^\star\gamma^\star}&=&  ie^2\varepsilon^{\mu\nu\rho\sigma}{q_1}_\mu{q_2}_\nu{\epsilon^*_1}_\rho{\epsilon^*_2}_\sigma\,
  \mF_{P\gamma^\star\gamma^\star}\left(q_1^2,q_2^2\right)\, ,
  \label{eq:TFF-def}
 \end{eqnarray}
where Bose symmetry implies $\mF_{P\gamma^\star\gamma^\star}(q_1^2,q_2^2)=\mF_{P\gamma^\star\gamma^\star}(q_2^2,q_1^2)$.\\

\begin{figure}[!t]
  \centering\includegraphics[scale=0.75]{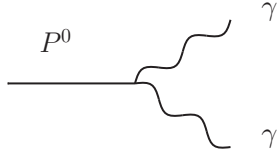}\caption{{\small Local contribution to the $P^0\to \gamma\gamma$ decay.}}\label{ChPT}
 \end{figure}

This amplitude can be divided into three kinds depending on the number of intermediate vector resonances contributing to each diagram. First we have contributions involving no resonances, as that shown in Figure \ref{ChPT}, which can be computed by means of the Wess-Zumino-Witten functional \cite{WZW} (of chiral order $p^4$) and the $\mathcal{O}(p^6)$ $\chi$PT Lagrangian \cite{Bijnens}. This gives
\begin{eqnarray}
  \mathcal{M}_{\text{Loc}}&=&i\varepsilon^{\mu\nu\rho\sigma}{q_1}_\mu{q_2}_\nu{\epsilon^*_1}_\rho{\epsilon^*_2}_\sigma\frac{2e^2}{3F}\times\nonumber\\
  &&\left\{-\left[
  \frac{N_C}{8\pi^2}+8(q_1^2+q_2^2)C_{22}^W\right]C_{P_0}+32C_7^WC_{P_7}+64C_8^WC_{P_8}\right\},\;\;
 \end{eqnarray}
where the constants $C_{P_0}$, $C_{P_7}$ and $C_{P_8}$ depend on the pseudo-Goldstone boson considered and are given in table~\ref{tab:Local_const}.

\begin{table}[!ht]
    \centering
    \begin{tabular}{c c c c}\hline\hline
         & $C_{P_0}$ & $C_{P_7}$ & $C_{P_8}$ \\\hline
    $\pi$ & 1 & $m_\pi^2$ & 0\\
    $\eta$& $\left({5C_q-\sqrt{2}C_s}\right)/{3}$ & $\left({5C_qm_\pi^2-\sqrt{2}C_s\Delta_{2K\pi}^2}\right)/{3}$ &
    $2C_qm_\pi^2-\sqrt{2}C_s\Delta_{2K\pi}^2$\\
    $\eta'$& $\left({5C_q'+\sqrt{2}C_s'}\right)/{3} $& $\left({5C_q'm_\pi^2+\sqrt{2}C_s'\Delta_{2K\pi}^2}\right)/{3}$&
    $2C_q'm_\pi^2+\sqrt{2}C_s'\Delta_{2K\pi}^2$\\\hline\hline
    \end{tabular}
    \caption{Values of $C_{P_i}$ for the local interaction.
    }
    \label{tab:Local_const}
\end{table}

\begin{figure}[!t]
  \centering\includegraphics[scale=0.75]{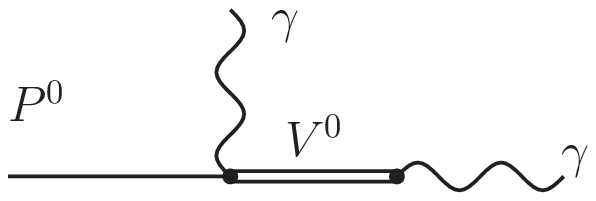}\caption{{\small Contribution to the $P^0\to \gamma\gamma$ decay with one resonance exchange.}}\label{1RR}
 \end{figure}

 The contributions involving one vector resonance exchange (fig.~\ref{1RR}) are
 \begin{eqnarray}
  \mathcal{M}_{1R}&=&-i\varepsilon^{\mu\nu\rho\sigma}{q_1}_\mu{q_2}_\nu{\epsilon_1^*}_\rho{\epsilon_2^*}_\sigma\frac{4\sqrt{2}e^2}{3M_VF}\times\\\nonumber
  &&\hspace*{-10ex}\sum_{i=\rho^0,\omega,\phi}C_{V_i}\left\{\frac{4c_3(C_{1R}^m)_i+(C_{1R}^d)_i\left[q_1\cdot q_2(c_1+c_2-c_5)+q_1^2(c_2-c_6)+q_2^2c_1\right]}{M_i^2-q_1^2}+(q_1\leftrightarrow q_2)\right\},
 \end{eqnarray}
where the constants $C_V$ are
\begin{eqnarray}
 C_{V_i}=\left\{\begin{array}{cc}\frac{1}{3}(F_V+8m_\pi^2\lambda_{V})&\text{for }\omega,\\(F_V+8m_\pi^2\lambda_{V})&\text{for }\rho^0,\\
 \frac{\sqrt{2}}{3}\left(F_V+8\Delta_{2K\pi}^2\lambda_{V}\right)&\text{for }\phi,\\ \end{array}\right.
\end{eqnarray}
and the constants $C_{1R}^d$ and $C_{2R}^m$ are given in table~\ref{tab:C1R}.

\begin{table}[!ht]
    \centering
\begin{tabular}{cccc}\hline\hline
            $C_{1R}^d$&$\omega$&$\rho^0$&$\phi$\\\hline
            $\pi^0$&3&1&0\\
            $\eta$&$C_q$&$3C_q$&$-2C_s$\\
            $\eta'$&$C_q'$&$3C_q'$&$2C_s'$\\\hline\hline
           \end{tabular}\hspace*{8ex}\begin{tabular}{cccc}\hline\hline
            $C_{1R}^m$&$\omega$&$\rho^0$&$\phi$\\\hline
            $\pi^0$&$3m_\pi^2$&$m_\pi^2$&0\\
            $\eta$&$m_\pi^2C_q$&$3m_\pi^2C_q$&$-2\Delta_{2K\pi}^2C_s$\\
            $\eta'$&$m_\pi^2C_q'$&$3m_\pi^2C_q'$&$2\Delta_{2K\pi}^2C_s'$\\\hline\hline
           \end{tabular}\caption{Values of $C_{1R}^d$ (left) and $C_{1R}^m$ (right) for the couplings between different mesons.}\label{tab:C1R}
\end{table}

\begin{figure}[!t]
  \centering\includegraphics[scale=0.75]{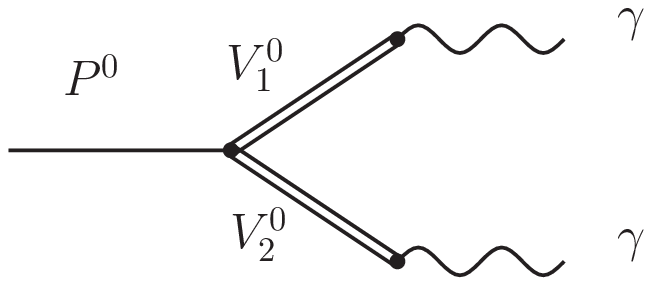}\caption{{\small Contribution to the $P^0\to \gamma\gamma$ decay with two resonances exchange.}}\label{2RR}
 \end{figure}

 The two-vector meson exchange contributions (fig.~\ref{2RR}) are given by
 \begin{eqnarray}
  \mathcal{M}_{2R}&=&i\varepsilon^{\mu\nu\rho\sigma}{q_1}_\mu{q_2}_\nu{\epsilon_1^*}_\rho{\epsilon_2^*}_\sigma\frac{2\sqrt{2}e^2}{F}\times\\\nonumber
  &&\sum_{i,j=\rho^0,\omega,\phi}C_{Vi}C_{Vj}\frac{8d_2(C_{2R}^m)_{ij}+(C_{2R}^d)_{ij}\left[2(d_1-d_3)q_1\cdot q_2+d_1(q_1^2+q_2^2)\right]}{(M_i^2-q_1^2)(M_j^2-q_2^2)},
\end{eqnarray}
where the values for the constants $C_{2R}^d$ and $C_{2R}^m$ can be read from table~\ref{tab:C2R}.
\begin{table}[!ht]
    \centering
\begin{tabular}{ccccc}\hline\hline
            $C_{2R}^d$&$\omega\omega$&$\rho^0\rho^0$&$\omega\rho^0$&$\phi\phi$\\\hline
            $\pi^0$&0&0&$\sqrt{2}$&0\\
            $\eta$&$\sqrt{2}C_q$&$\sqrt{2}C_q$&0&$-2C_s$\\
            $\eta'$&$\sqrt{2}C_q'$&$\sqrt{2}C_q'$&0&$2C_s'$\\\hline\hline
           \end{tabular}\\\hspace*{1ex}\\
           \vspace*{0.2cm}
           \begin{tabular}{ccccc}\hline\hline
            $C_{2R}^m$&$\omega\omega$&$\rho^0\rho^0$&$\omega\rho^0$&$\phi\phi$\\\hline
            $\pi^0$&0&0&$\sqrt{2}m_\pi^2$&0\\
            $\eta$&$\sqrt{2}m_\pi^2C_q$&$\sqrt{2}m_\pi^2C_q$&0&$-2\Delta_{2K\pi}^2C_s$\\
            $\eta'$&$\sqrt{2}m_\pi^2C_q'$&$\sqrt{2}m_\pi^2C_q'$&0&$2\Delta_{2K\pi}^2C_s'$\\\hline\hline
           \end{tabular}\caption{Values of $C_{2R}^m$ (up) and $C_{2R}^m$ (bottom) for the couplings between different mesons.}\label{tab:C2R}
\end{table}

Adding up the various $\pi^0\to\gamma^\star\gamma^\star$ contributions we obtain the $\pi^0$-TFF,
  \begin{eqnarray}\label{FormF}
  \mathcal{F}_{\pi^0\gamma^\star\gamma^\star}(q_1^2,q_2^2)&=&\frac{2}{3F_\pi}\left\{-\frac{N_C}{8\pi^2}+32m_\pi^2C_7^W+\left[
  -8q_1^2{C_{22}^W}\frac{}{}\right.\right.\nonumber\\&&
  +\frac{2(F_V+8m_\pi^2\lambda_V)^2\left(d_{3}(q_1^2+q_2^2)+d_{123}m_\pi^2\right)}{D_\rho(q_1^2)D_\omega(q_2^2)}\nonumber\\&&
  -\frac{\sqrt{2}(F_V+8m_\pi^2\lambda_V)}{M_V}\left(m_\pi^2c_{1235}-q_1^2c_{1256}+
  q_2^2c_{125}\right)\times\nonumber\\&&
  \left(\frac{1}{D_\rho(q_1^2)}+\frac{1}{D_\omega(q_1^2)}\right)
  \left.\left.\frac{}{}+\left(q_1\leftrightarrow q_2\right)\right]^\frac{}{}\right\},
 \end{eqnarray}
where $D_R(s)=M_R^2-s$. Note that the pion wave-function renormalization changes the global $F^{-1}$ factor to $F_\pi^{-1}$ in the large--$N_C$ limit~\cite{Bernard:1991zc,SanzCillero:2004sk,Guo:2014yva}.  \\

For simplicity, in addition to the couplings $c_3$ and $d_2$, we will use the following combinations of constants:
 \begin{subequations}
 \begin{align}
     c_{1235}&=c_1+c_2+8c_3-c_5,\\
     c_{1256}&=c_1-c_2-c_5+2c_6,\\
     c_{125}&=c_1-c_2+c_5,\\
     d_{123}&=d_1+8d_2-d_3.
 \end{align}
 \end{subequations}
 For the $\eta$ meson we find the TFF,
   \begin{eqnarray}
  \mathcal{F}_{\eta\gamma^\star\gamma^\star}(q_1^2,q_2^2)&=&\frac{2}{3F}\left\{-\frac{\left(5 C_q-\sqrt{2} C_s\right) }{3}\left[8(q_1^2+q_2^2)C_{22}^W
  +\frac{N_C}{8\pi^2}\right]\right.
  \nonumber\\
  && \hspace*{-1.75cm}
  + 32C_7^W\frac{5C_qm_\pi^2-\sqrt{2}C_s\Delta_{2K\pi}^2}{3}+64C_8^W\left(2C_qm_\pi^2-\sqrt{2}C_s\Delta_{2K\pi}^2\right)
  \nonumber\\
  &&\hspace*{-1.75cm}
  +\left[-\frac{\sqrt{2}C_q}{3M_V}\frac{(F_V+8m_\pi^2\lambda_V)\left(-8c_3\Delta_{\eta\pi}^2+c_{1235}m_\eta^2-c_{1256}q_1^2+c_{125}q_2^2\right)}{D_\rho(q_1^2)}
  \right.\nonumber\\
  &&\hspace*{-1.75cm}
  -\frac{3\sqrt{2}C_q}{M_V}\frac{(F_V+8m_\pi^2\lambda_V)\left(-8c_3\Delta_{\eta\pi}^2+c_{1235}m_\eta^2-c_{1256}q_1^2+c_{125}q_2^2\right)}{D_\omega(q_1^2)}
  \nonumber\\
  &&\hspace*{-1.75cm}
  +\frac{4C_s}{3M_V}\frac{(F_V+8\Delta_{2K\pi}^2\lambda_V)\left(c_{1235}m_\eta^2-c_{1256}q_1^2+c_{125}q_2^2+8c_3\Delta_{2K\pi\eta}^2\right)}{D_\phi(q_1^2)}
  \nonumber\\
  &&\hspace*{-1.75cm}
   +\frac{3 C_q
   (F_V+8m_\pi^2\lambda_V)^2 \left(-8 d_2 \Delta_{\eta\pi}^2+d_{123} m_{\eta
   }^2+d_3 (q_1^2+ q_2^2)\right) }{D_{\rho }\left(q_1^2\right)
   D_{\rho }\left(q_2^2\right)}\nonumber\\
   &&\hspace*{-1.75cm}
   +\frac{C_q
   (F_V+8m_\pi^2\lambda_V)^2 \left(-8 d_2 \Delta_{\eta\pi}^2+d_{123} m_{\eta
   }^2+d_3 (q_1^2+ q_2^2)\right)}{3 D_{\omega
   }\left(q_1^2\right) D_{\omega }\left(q_2^2\right)}\nonumber\\
   &&\hspace*{-1.75cm}
    -\frac{2 \sqrt{2} C_s (F_V+8\Delta_{2K\pi}^2\lambda_V)^2 \left(d_{123}
    m_{\eta }^2+d_3( q_1^2+ q_2^2)+8 d_2\Delta_{2K\pi\eta }^2
    \right)}{3 D_{\phi }\left(q_1^2\right)
    D_{\phi }\left(q_2^2\right)}\nonumber\\
    &&
   \left.\left.\frac{}{}+(q_1\leftrightarrow q_2)^\frac{}{}\right]^\frac{}{}\right\},\label{etaTFF}
 \end{eqnarray}
where $\Delta_{\eta\pi}^2=m_\eta^2-m_\pi^2$
and $\Delta_{2K\pi\eta }^2=2 m_K^2-m_{\pi }^2-m_{\eta }^2$. The result for the $\eta'$ can be obtained from the previous one by substituting $C_q\to C_q'$, $C_s\to -C_s'$ and $m_\eta\to m_{\eta'}$.
Notice that by taking the chiral limit one recovers the expressions given previously in refs.~\cite{Kampf,us}.

We would like to remark that the value of pseudoscalar decay constant in the chiral limit, $F$, is not required in our R$\chi$T description. In the $\pi^0$-TFF it combines into the global $F_\pi^{-1}$ factor in eq.~(\ref{FormF}),
with $F_\pi=92.2$ MeV \cite{PDG}.
In the same way, the $\eta$ and $\eta'$ form-factors only depend on the combinations
$C_q/F$, $C_s/F$, $C_q'/F$ and $C_s'/F$ which, in turn, only depend on the mixing angles $\theta_8$, $\theta_0$ and the value of $f_8$ and $f_0$ (not on their ratios $f_8/F$ and $f_0/F$).

 Since we want to make use of the $VVP$ Green's function short-distance  constraints~\cite{Kampf},
 the analysis is incomplete if the pseudoscalar resonance multiplet, $P'$, --included in ref.~\cite{Kampf}--  is not considered. The $d_m$ operator introduces a mixing between $P'$ and the pseudo-Goldstone states $P$ proportional to the quark masses. Thus, the contributions from the heavy pseudoscalar nonet to the $P$--TFFs begin at $\cO(m_P^2)$ and, at that order,
 take the form
\begin{eqnarray}
  \mathcal{F}_{P\to P^{'\star}\to\gamma^\star\gamma^\star}&=&
  C_{P_7}\, \times \,\Frac{ 2\sqrt{2}d_m }{F M_{P'}^2} \, \mathcal{F}_{P_{I=1}^{'\star}(p)\to\gamma^\star\gamma^\star}\bigg|_{p^2\to 0} \, ,
 \label{eq:Pprime-contribution}
 \end{eqnarray}
in terms of the $P^{\prime\star}(p)\to\gamma^\star(q_1)\gamma^\star(q_2)$ transition amplitude in the chiral and large--$N_C$ limits, provided here
for the isospin $I=1$ resonance by
 \begin{eqnarray}
  \mathcal{F}_{P_{I=1}^{'\star}(p)\to\gamma^\star\gamma^\star}&=&
   \frac{8\sqrt{2}}{3}\left[4\kappa_5^P-\sqrt{2}\kappa_3^{PV}F_V\left(\frac{1}{M_V^2 - q_1^2} + \frac{1}{M_V^2 - q_2^2}\right)\right.\nonumber\\&&
  \qquad\qquad \left.+\frac{\kappa^{PVV}F_V^2}{(M_V^2 - q_1^2)(M_V^2 - q_2^2)}\right]\, ,
 \label{eq:Pprime-TFF}
 \end{eqnarray}
 where $C_{P_7}=\cO(m_P^2)$ and $M_{P'}$ is the $P'$ mass in the chiral and large--$N_C$ limits.
 $U(3)$ breaking effects in the $P'$ multiplet would contribute to our $P$--TFFs at higher orders in $m_P^2$.
 This result for the off-shell $P^{'\star}\gamma^\star\gamma^\star$ transition is in agreement with ref.~\cite{Kampf} and the $U(3)$ limit of eq.~(\ref{eq:Pprime-contribution}) (given by $m_\pi=m_K$ in $C_{P_7}$) coincides with $U(3)$ symmetric contribution to the $P$--TFFs in ref.~\cite{us}.
  Comparing these expressions for the $P'$ contributions with those in eqs.~(\ref{FormF}) and~(\ref{etaTFF}), it becomes evident that these contributions can be included merely by a redefinition of three parameters, namely, $C_7^W$, $c_{3}$, $d_2$:~\footnote{
  One does not need to take into account the corrections from $\lambda_V$ and $e_m^V$ in eq.~(\ref{eq:Pprime-TFF}) for these redefinitions, as we are only interested in the $P'$ contributions to the $P$--TFFs at $\cO(m_P^2)$. }
 \begin{subequations}
 \begin{align}
 C_7^W\,\, \longrightarrow \,\,  {C_7^W}^\star&=C_7^W + \frac{2d_m\kappa_5^P}{M_{P'}^2}\, ,
 \\
 c_{3}\,\, \longrightarrow \,\,  \,\,  \,\,  c_{3}^\star\,\, &=c_{3} +\frac{d_mM_V\kappa_3^{PV}}{M_{P'}^2}\, ,
 \\
 d_{2}\,\, \longrightarrow \,\,  \,\,  \,\,   d_{2}^\star\,\, & = d_{2} + \frac{d_m \kappa^{PVV}}{2M_{P'}^2} \,.
 \label{eq:d2star}
 \end{align}
 \end{subequations}
Note that the R$\chi$T couplings $c_{1,2,5,6}$ and $d_{1,3}$ do not need to be corrected.
However, in order to take the $P'$ contributions into account, we will also need to shift our auxiliary combinations
\begin{subequations}
\begin{align}
c_{1235}\,\, \longrightarrow \,\,  & c_{1235}^\star=
c_1+c_2+8c_3^\star - c_5\, ,
 \\
 d_{123}\,\, \longrightarrow \,\, & d_{123}^\star = d_1+8d_2^\star-d_3\,.
 \end{align}
 \end{subequations}

 \section{Short Distance Constraints}\label{SD}

 We will constrain the R$\chi$T parameters by imposing on the $\mF_{P\gamma^\star\gamma^\star}$ TFFs ($P=\pi^0,\eta,\eta'$) the high energy behaviour prescribed by QCD~\cite{BL,Nesterenko}:
 \bear
\mathcal{F}_{P\gamma^\star\gamma^\star}(q^2,q^2)\,\, \stackrel{Q^2\to\infty}{\longrightarrow}\,\, 0\, ,
\qquad\qquad\
\mathcal{F}_{P\gamma^\star\gamma^\star}(q^2,0)\,\,  \stackrel{Q^2\to\infty}{\longrightarrow}\,\,  0\, ,
 \eear
with $Q^2=-q^2$.
These conditions are applied order by order in the $m_P^2$ expansion, separately: first in the massless limit, at $\cO(m_P^0)$, and then at $\cO(m_P^2)$.
We emphasize that in this work we are not matching the precise QCD predictions for the coefficients of
the $1/Q^2$ expansion~\cite{BL,Nesterenko},
just requiring the vanishing of the TFF at high momentum transfer.
We will see that these results are found to be compatible with previous results~\cite{Kampf,SDConst} from the analysis of the VVP Green's function and other observables.
The constraints from $\cO(m_P^4)$ or higher are not taken into account, as we did not considered R$\chi$T operators contributing at that order. \\

The high-energy analysis of the TFF leads to the relations:
\begin{itemize}
\item{\bf $\pi^0$-TFF, $\hspace*{1ex}\cO(m_P^0)$ :}
 \begin{subequations}
  \begin{align}
   C_{22}^W=&\, 0,\\
   c_{125}=&\, 0,
   \label{eq:c125}\\
   c_{1256}=&-\frac{N_CM_V}{32\sqrt{2}\pi^2F_V},\\
   d_3=&-\frac{N_CM_V^2}{64\pi^2F_V^2}.
  \end{align}
 \end{subequations}
\item{\bf $\pi^0$-TFF, $\hspace*{1ex}\cO(m_P^2)$ :}
 \begin{subequations}
  \begin{align}
   \lambda_V=&-\frac{32\pi^2F_V}{N_C}C_7^{W
   \star}\, , \\
    c_{1235}^\star =& \frac{N_CM_Ve_m^V}{8\sqrt{2}\pi^2F_V}
    +\Frac{ N_C M_V^3 \lambda_V}{4\sqrt{2}\pi^2 F_V^2}\, .
  \end{align}
 \end{subequations}
\item{\bf Additional $\eta$-TFF constraints, $\cO(m_P^2)$:}
\begin{subequations}
  \begin{align}
   C_{8}^W=&\, 0,\\
   c_3^{\star} =&\frac{c_{1235}^\star}{8}=\frac{N_CM_Ve_m^V}{64\sqrt{2}\pi^2F_V}
    +\Frac{ N_C M_V^3 \lambda_V}{32\sqrt{2}\pi^2 F_V^2}\, .
  \end{align}
 \end{subequations}
Notice that the implicit $\kappa_3^{PV}$ cancels out in the last equation so it should be actually read as $c_3=c_{1235}/8$. This, in combination with the constraint~(\ref{eq:c125}), yields $c_1=c_2-c_5=0$.
\item{\bf No additional $\eta'$-TFF constraints:}
up to $\cO(m_P^2)$, the analysis of the $\eta'$-TFF casts the same relations provided by the $\pi^0$ and $\eta$ form-factors.
\end{itemize}
We note that we did not need to perform the $m_P^2$ expansion of the mixing coefficients $C_{q/s}^{(')}$. One obtains the previous set of consistent relations for any value of $C_{q/s}^{(')}$: our conditions are actually requiring a good high-energy behaviour to the bare $\phi^a\to\gamma^\star\gamma^\star$ TFFs, which is inherited by the physical $\pi^0$, $\eta$ and $\eta'$ TFFs after taking into account the field renormalizations and mixings provided by $C_\pi$ and $C_{q/s}^{(')}$.  \\

This allows us to by-pass the cumbersome problem of the large--$N_C$ limit in the $\eta-\eta'$ mixing~\cite{Kaiser:2000gs}, where $1/N_C$ and quark mass corrections are found to be of a similar numerical order~\cite{Kaiser:2000gs,eta-etap,Guo:2015xva}. \\

We will supplement these TFF constraints with the additional conditions
derived from the short-distance analysis of the VVP Green's function $\Pi(p^2,q^2,r^2)$~\cite{Kampf}. By matching the leading terms of the QCD OPE at $p^2,q^2,r^2\to\infty$ within the chiral and large--$N_C$ limits, ref.~\cite{Kampf} obtains~\footnote{In this article we use the notation from ref.~\cite{VVP}. It relates to the Lagrangian~\cite{Kampf} through $c_{1235}=M_V(2\kappa_{12}^V+8\kappa_{14}^V+\kappa_{16}^V)$, $c_{1256}=M_V(2\kappa_{12}^V+\kappa_{16}^V)$, $c_{125}=M_V(2\kappa_{12}^V+\kappa_{16}^V-2 \kappa_{17}^V)$,
$c_3=M_V(\kappa_{13}^V+\kappa_{14}^V)$, $d_{123}=8\kappa_2^{VV}-\kappa_3^{VV}$
and $d_3=\kappa_3^{VV}$~\cite{SDConst}.},
\begin{itemize}
\item{\bf $VVP$ Green's function, $\hspace*{1ex}\cO(m_P^0)$ :}
\bear
&&\hspace*{-1cm} c_{125}\,=\, c_{1235}\,=\, 0\,, \qquad c_{1256}= -\Frac{N_C M_V}{32\sqrt{2}\pi^2 F_V}\, ,
\nn\\
&&\hspace*{-1cm} d_3 \,=\, -\Frac{N_C M_V^2}{64\pi^2 F_V^2 } + \Frac{F^2}{8F_V^2} + \Frac{4\sqrt{2}d_m\kappa_3^{PV}}{F_V}\, , \qquad
d_{123}\,=\, \Frac{F^2}{8 F_V^2}\, ,
\qquad \kappa_5^P\,=\,0\, ,
\nn\\
&&\hspace*{-1cm} C_7^W\,=\,C_8^W\, =\,C_{22}^W \,=\, 0\, .
\eear
\end{itemize}
In the last line, we added the VVP constraints for the R$\chi$T couplings $C_{j}^W$ in $\mL_{\rm non-R}$.
It is remarkable that these relations are compatible with our TFF relations above, being the relations for $c_{125}$, $c_{1256}$, $C_8^W$ and $C_{22}^W$ identical.
Combining these VVP constraints~\cite{Kampf} with the previous ones from the high-energy TFF analysis up to $\cO(m_P^2)$, we conclude the constraints
\bear
&&c_1\,=\, c_2-c_5\, =\, c_3\,=\,c_{125}\,=\, c_{1235}\,=\,  0\,,
\nn\\
&&c_{1235}^\star\,=\, 8 c_3^\star \,=\, \, \Frac{8 d_m \kappa_3^{PV} M_V}{M_{P'}^2}\,=\, \Frac{N_C e_m^V M_V}{8\sqrt{2} \pi^2 F_V}\, ,
\nn\\
&& d_{123}^\star\,=\, \Frac{F^2}{8 F_V^2} + \Frac{4 d_m\kappa^{PVV}}{M_{P'}^2}\, ,
 \qquad\qquad d_{123}\,=\, \Frac{F^2}{8 F_V^2}\, ,
\nn\\
&& d_3=-\frac{N_CM_V^2}{64\pi^2F_V^2}\,,\qquad\qquad c_{1256}= -\Frac{N_C M_V}{32\sqrt{2}\pi^2 F_V}\,,\nn\\
&&{C_7^W}^\star\,=\, \lambda_V=0 \,,
 \qquad \kappa_5^P=0\, , \qquad C_7^W={C_8^W}=C_{22}^W=0.
\label{eq:new-SD-rels}
\eear
   The relations obtained in this way fix all the parameters in the form-factors except for $d_{123}^\star$, $d_2^\star$, $M_V$, $e_m^V$
   and the four $\eta-\eta'$ mixing parameters. Particularly, among the operators violating $U(3)$ explicitly, those with coefficients $c_3$, ${C_7^W}^\star$, $C_8^W$ and $\lambda_V$ vanish, according to eq.~(\ref{eq:new-SD-rels}).
   Although there is a relation for $d_{123}$ from the OPE study of the VVP Green's function,
   $d_{123}^\star$ is not well known since the $\kappa^{PVV}$ coupling remains unconstrained~\cite{Kampf}. The unrestricted parameters are fitted in this work using $\mathcal{F}_{P\gamma^\star\gamma^\star}(q^2,0)$ experimental data for $\pi^0$, $\eta$ and $\eta'$. The results are given in the next section.
   There is one more constraint for $\kappa_3^{PV}$ from the combination of the TFF and VVP relations for $d_3$
   in the chiral and large--$N_C$ limits~\cite{Kampf}, which, however will not be considered in detail in this work:
   \bear
   8 d_m \kappa_3^{PV}\,=\, -\Frac{F^2}{4\sqrt{2} F_V}\, <\, 0\, .
   \eear
   In principle, this, together with the previous constraints, leads to a negative vector mass splitting parameter $e_m^V$, in agreement with previous phenomenology~\cite{MassSplitting} and our later fits. This prediction $e_m^V=-2\pi^2 F^2/(N_C M_{P'}^2)$ is, however, one order of magnitude smaller for $F\sim F_\pi$ and $M_{P'}\sim 1.3$~GeV than the values required by the phenomenology.
   Nonetheless, one should be very cautious with the theoretical determinations of the $P'$ couplings: under the lightest $P'$ and $V$ resonance multiplet approximation it is possible to obtain a good high-energy behaviour for the $\pi^0$--TFF and VVP Green's function in the chiral limit but at the price of a $P'\to\gamma^\star\gamma$ TFF~(\ref{eq:Pprime-TFF}) which does not vanish a large momentum transfer. For this reason we have decided
   to leave $\kappa_3^{PV}$ --or equivalently $e_m^V$-- as a phenomenological parameter in our later fits, postponing further studies on the structure of the $P'$ Lagrangian for a future work.\\

 After using the SD constraints, the $\pi^0$-TFF results simplified into the form
 \begin{equation}\label{SimppiFF}
  \mathcal{F}_{\pi\gamma^\star\gamma^\star}(q_1^2,q_2^2)=\frac{32\pi^2m_\pi^2F_V^2d_{123}^\star -  N_C M_V^2M_\rho^2
   }{12 \pi ^2 F_\pi D_\rho(q_1^2)
   D_\rho(q_2^2)},
 \end{equation}
 Also the $\eta$-TFF is simplified to
 \begin{eqnarray}\label{SimpetaFF}
  \mathcal{F}_{\eta\gamma^\star\gamma^\star}(q_1^2,q_2^2)&=&\frac{1}{12\pi^2F D_\rho(q_1^2) D_\rho(q_2^2)D_\phi(q_1^2) D_\phi(q_2^2)}\times
  \\
   &&
   \hspace*{-2cm}\left\{-\frac{N_C M_V^2}{3     }\left[5C_q M_\rho^2D_\phi(q_1^2) D_\phi(q_2^2) - \sqrt{2}C_s M_\phi^2 D_\rho(q_1^2) D_\rho(q_2^2)\right]\right.
   \nonumber\\
   &&
   \hspace*{-2cm}+\frac{32\pi^2F_V^2d_{123}^\star m_\eta^2}{3}\left[(5C_q D_\phi(q_1^2) D_\phi(q_2^2)-\sqrt{2}C_s D_\rho(q_1^2) D_\rho(q_2^2)\right]
   \nonumber\\
   &&
   \hspace*{-2cm}\left.-\frac{256\pi^2F_V^2d_{2}^\star}{3}\left[(5C_q\Delta_{\eta\pi}^2 D_\phi(q_1^2) D_\phi(q_2^2)+\sqrt{2}C_s\Delta_{2K\pi\eta}^2 D_\rho(q_1^2) D_\rho(q_2^2)\right]\right\}.\nonumber
 \end{eqnarray}
The constrained $\eta'$-TFF is given by this expression with the same replacements indicated before after eq.~(\ref{etaTFF}): $C_q\to C_{q}'$, $C_s\to-C_s'$ and $m_\eta\to m_{\eta'}$.  \\

In the chiral and large--$N_C$ limits one has $m_P\to 0$ and we recover the previous result from ref.~\cite{Knecht}:
\bear
\mF_{\pi^0\gamma^\star\gamma^\star}(q_1^2,q_2^2) &=&
\, -\,  \Frac{ N_C M_V^4}{12 \pi ^2 F (M_V^2-q_1^2) (M_V^2-q_2^2)}\, , 
\label{eq:1R-chiral-TFF}
\eear
and $\mF_{\eta\gamma^\star\gamma^\star}(q_1^2,q_2^2) = \mF_{\pi^0\gamma^\star\gamma^\star}(q_1^2,q_2^2) \times (5 C_q -\sqrt{2} C_s)/3$. The result for the $\eta'$ TFF is analogous with $C_q\to C_{q'}$ and $C_s\to -C_{s'}$~\cite{us}~\footnote{
The values of $C_{q/s}^{(')}$ will depend on whether we take first the chiral or the large--$N_C$ limit, {\it i.e}, on the hierarchy of the associated scales. Nonetheless, in the $U(3)$ symmetric limit $m_q,N_C^{-1}\to 0$ (also for the $a_\mu^{HLbL}$ integration kernels), the sum of the $\eta$ and $\eta'$ contributions to the anomalous magnetic moment is proportional $[(5 C_q -\sqrt{2} C_s)/3]^2+ [(5 C_q' +\sqrt{2} C_s')/3]^2 = F^2 [f_8^{-2} + 8 f_0^{-2} + 4\sqrt{2} f_8^{-1}f_0^{-1} \sin{(\theta_8-\theta_0)}]/[3\cos(\theta_8-\theta_0)]= 3$, regardless of the order in which the limits are taken.
}. We find again at $\cO(m_P^2)$ the issue pointed out in previous studies of the TFF with only one vector multiplet at $\cO(m_P^0)$~\cite{Knecht}: it is not possible to match the $1/q^2$ coefficient prescribed by QCD at deep virtual momentum $Q^2\to\infty$ at the same time for both kinematical configurations $\mF_{P\gamma^\star\gamma^\star}(q^2,0)\approx 2 F/q^2$~\cite{BL} and $\mF_{P\gamma^\star\gamma^\star}(q^2,q^2)\approx 2 F/(3q^2)$~\cite{Nesterenko}. We will see that, even though the first limit is not imposed, our TFFs fitted to data approximately recover the asymptotic behaviour $2 F/q^2$. On the other hand, for the doubly off-shell TFFs we always find $\mF_{P\gamma^\star\gamma^\star}(q^2,q^2)
\stackrel{Q^2\to\infty}{=} \cO(1/q^4)$. 
{This deficiency of our description might lead to a slight
underestimation of our result for $a_\mu^{P,HLbL}$ that will be later discussed.}  \\

We would like to remark that the $P\gamma^\star\gamma^\star$ and $VVP$ short-distance relations provided here are fully compatible with the constraints from $\tau\to P^-\gamma\nu_\tau$~\cite{Guo:2010dv} and $\tau\to(PV)^-\nu_\tau$~\cite{Guo:2008sh}. The high-energy conditions from $\tau\to (KK\pi)^- \nu_\tau$~\cite{Dumm:2009kj} and $\tau\to \eta \pi^- \pi^0 \nu_\tau$~\cite{Dumm:2012vb} are also compatible with the relations in the present article provided $F_V=\sqrt{3}F$~\cite{SDConst}. Nevertheless, the outcomes in this article will not rely on this last relation, even though it will be useful for comparison. \\

\section{{Form factor experimental data analysis} }\label{Fits}
\subsection{Fit to experimental TFF}

After demanding the high-energy TFF and $VVP$ conditions, these form-factors depend only on the vector mass parameters $M_V$ and $e_m^V$, the combinations $F_V^2 d_{123}^\star$ and, in the $\eta$ and $\eta'$ cases,  $F_V^2 d_2^\star$ and the coefficients
$C_{q/s}^{(')}/F$
(which only depend on the mixing parameters $\theta_8$, $\theta_0$, $f_8$ and $f_0$)~\footnote{Instead of $F_V^2 d_{123}^\star$ and $F_V^2 d_2^\star$ we could have chosen $\kappa^{PVV}$ and $d_2$ for our set of independent fit parameters, related through eqs.~(\ref{eq:d2star}) and~(\ref{eq:new-SD-rels}). However, we have preferred to use the former ones for sake of clarity.}. All fitted parameters (but $M_V$) break explicitly the $U(3)$ flavor symmetry. \\

We will determine these 8 parameters
by means of a fit to the available space-like experimental data.
Notice, however, that our fit will be completely insensitive to $F_V$, since it only appears in the form-factors multiplying $d_2^\star$ and $d_{123}^\star$.
 Therefore, for convenience and to ease the comparison we
 will show the results for the combinations
 \bear
 \bar{d}_2 \equiv \Frac{F_V^2 d_2^\star}{3F_\pi^2}\, ,
 \qquad
 \bar{d}_{123} \equiv \Frac{F_V^2 d_{123}^\star}{3F_\pi^2}\, ,
 \eear
with $F_\pi=92.2$~MeV~\cite{PDG}.\\

Since we are mainly focused on the application of the transition Form Factor in the pseudoscalar exchange contribution to the $a_\mu^{HLbL}$, we will only perform fits to data with negative squared photon momenta, $q^2\le0$. There are two reasons for this, the first one is that being a NLO effect in the $1/N_C$ expansion, the width of the resonances is needed to obtain a finite result in the $q^2>0$ region of momenta, and to be consistent, we would need to consider also all NLO terms in the $1/N_C$ expansion; the other one is that the observables in the time-like region may get large contributions from disregarded photon radiative corrections \cite{NLODD}.\\

The $\Gamma(P\to\gamma\gamma)$ decay widths give relevant information of the form-factor at very low energies and help reduce the error in the $\eta-\eta'$ mixing parameters. The values for the $\Gamma(P\to\gamma\gamma)$ decays are taken from the Particle Data Group (PDG)~\cite{PDG}. The relation between the form-factor and the on-shell photons decay width is given by
 \begin{equation}
     |\mathcal{F}_{P\gamma^\star\gamma^\star}(0,0)|^2=\frac{64\pi}{(4\pi\alpha)^2}\frac{\Gamma(P\to\gamma\gamma)}{m_P^3}.
 \end{equation}

 We use CLEO \cite{CLEO}, CELLO \cite{CELLO}, LEP \cite{Lep}, BaBar \cite{BaBar,BABAR:2011ad} and Belle \cite{Belle} data of the form-factors along with the decay width to real photons. Since the PDG takes into account LEP data at $q^2=0$ for the $\eta'\to\gamma\gamma$ transition, we have removed this bin to avoid double counting. We fit our form-factors for $\pi^0$, $\eta$ and $\eta'$ to all these data, simultaneously.\\

 In order to stabilize the fit, we have added a contribution to the $\chi^2$ given by the values of the parameters $\theta_{8/0}$ and $f_{8/0}$ from previous determinations~\cite{Kaiser:2000gs,eta-etap,mixing}, namely
 \begin{subequations}\label{stabilizer}
 \begin{align}
     \theta_8 =& (-21.2\pm1.6)^\circ,
     \label{eq:mix-input1}\\
     \theta_0 =& (-9.2\pm1.7)^\circ,
     \label{eq:mix-input2}\\
     f_8 =& (1.26\pm0.04)F_\pi=(116.2\pm3.7) \mathrm{\hspace*{1ex}MeV},
     \label{eq:mix-input3}\\
     f_0 =& (1.17\pm0.03)F_\pi=(107.9\pm2.8) \mathrm{\hspace*{1ex}MeV}.
     \label{eq:mix-input4}
 \end{align}
 \end{subequations}
 This gives a very small contribution $\Delta\chi^2\approx 1.5$. These inputs could be understood as a prior distribution for the mixing parameters given by~\cite{Kaiser:2000gs,eta-etap,mixing}.
 If one, however, takes the numbers given in ref.~\cite{Guo:2015xva}~\footnote{
 $U(3)$ large--$N_C$ $\chi$PT, NNLO Fit-B to lattice data~\cite{Guo:2015xva}:
 $\theta_8=(-27.9\pm 1.0\pm 1.4)^\circ$, $\theta_0=(-6.8\pm 0.9 \pm 3.7)^\circ$, $f_8=(126.5\pm 1.2\pm 11.8)$~MeV and
 $f_0=(109.1\pm 1.3\pm 5.9)$~MeV. },
 $f_0$ and $\theta_0$ coincide with the results of our best fit in table~\ref{tab:Fitted param},
 but we get $\theta_8=(-23.8\pm0.9)^\circ$ and $f_8=(151\pm9)$ MeV, which is no longer compatible with the results from table \ref{tab:Fitted param}. Also, since $e_m^V$ and $M_V$ are correlated to the mixing parameters, these values change to $M_V=(777\pm10)$ MeV and $e_m^V=(-51\pm17)\cdot10^{-2}$. This input rises the $\chi^2$ by $\sim12$. Despite the different values of these parameters, we get a value for $a_\mu^{P,HLbL}$ still compatible with the result from our best fit (see section \ref{Concl}). If one uses $F$ instead of $F_\pi$ in eqs.~(\ref{eq:mix-input3}) and~(\ref{eq:mix-input4}), the values for the parameters are completely compatible with those from our best fit. With this input we also get a result compatible with our best fit $a_\mu^{P,HLbL}$ (see secs.~\ref{sec:g-2} and~\ref{Concl}). If, furthermore, we remove the stabilizing conditions, eqs.~(\ref{eq:mix-input1})--(\ref{eq:mix-input4}), we end up with values for the mixing parameters that are incompatible with previous determinations \cite{Kaiser:2000gs,eta-etap,mixing,Guo:2015xva}:
 $\theta_8=(-27\pm 4)^\circ$,
 $\theta_0=(-5\pm 8)^\circ$,
 $f_8=(220\pm 40)$~MeV and $f_0=(110\pm 30)$~MeV
 (though this leads to $a_\mu^{P,HLbL}=(8.81\pm 0.16)\cdot 10^{-10}$, compatible with our best fit in secs.~\ref{sec:g-2} and~\ref{Concl}).
 The reason for this is that, as can be seen in table \ref{tab:CorrWoB} 
 (and tables \ref{tab:CorrWB} and \ref{tab:CorrMV}), there is a large
 correlation between $\mP_1$ and $\theta_8$ and between $\mP_2$ and $\theta_0$ and $f_0$ 
 (in addition to the strong correlations between the four mixing parameters)~\footnote{  
 {This correlations are even stronger if the contribution from eq.~(\ref{stabilizer}) is removed from the fit (excluding BaBar $\pi^0$--TFF data): 
 $\theta_8$, $\theta_0$ and $f_0$ are highly correlated with $\mP_2$ (with correlations 0.882, -0.887 and -0.914, respectively); 
 $f_8$ also shows strong correlations with $\mP_1$ (0.469), $M_V$ (-0.474) and $e_m^V$ (-0.440).  }  
 }. 
 Therefore, the fit to the TFF data alone is not really sensitive to the four mixing parameters but to combinations of them and $\mP_{1,2}$. 
 The stabilizing conditions in eq. (\ref{stabilizer})
 lessen the effect of such high correlations avoiding, thus, a non-physical region for the mixing parameters.
 \\

 \begin{table}[!t]
    \centering
    \begin{tabular}{cccc}\hline
        & & (Best Fit) & \\
        & `fit 1' & `fit 2' & `fit 3' \\
        & With $\pi^0$-BaBar & Without $\pi^0$-BaBar&Fixing $M_V$ and $e_m^V$
        \\\hline\hline \vspace*{-0.25cm}\\
    $\mP_1$
    & $-0.2\pm 1.0$
    & $0.0\pm 1.0$
    & $0.0\pm 1.0$
    \\
    $\mP_2$
    & $0.5\pm 1.0$
    & $0.0\pm0.5$
    & $0.0\pm 1.0$
    \\
     $\bar{d}_2$
     & $(-2.9\pm1.7)\cdot10^{-2}$
     & $(-2.7\pm1.7)\cdot10^{-2}$
     & $(-3\pm2)\cdot10^{-2}$
     \\
     $\bar{d}_{123}$
     & $(-2.5\pm1.5)\cdot10^{-1}$
     & $(-2.3\pm1.5)\cdot10^{-1}$
     & $(-3\pm2)\cdot10^{-1}$
     \\
     $M_V$
     & $(799\pm5)$ MeV
     & $(791\pm6)$ MeV
     & 764.3 MeV $^\dagger$
     \\
     $e_m^V$
     & $-0.35\pm0.10$
     & $-0.36\pm0.10$
     & $-0.228$ $^\dagger$
     \\
     $\theta_8$
     & $(-19.5\pm0.9)^\circ$
     & $(-19.5\pm0.9)^\circ$
     & $(-21.7\pm0.9)^\circ$
     \\
     $\theta_0$
     & $(-9.5\pm1.6)^\circ$
     & $(-9.5\pm1.6)^\circ$
     & $(-10.4\pm1.6)^\circ$
     \\
     $f_8$
     & $(118\pm4)$ MeV
     & $(118\pm3)$ MeV
     & $(118\pm3)$ MeV
     \\
     $f_0$
     & $(108\pm3)$ MeV
     & $(107.5\pm 1.0)$ MeV
     & $(107\pm3)$ MeV
     \\\hline
     $\chi^2$/dof & $ 150./101$ & $ 69./84$ & $101./86$\\\hline\hline
    \end{tabular}
    \caption{{\small Values for the fitted parameters with (first column) and without (second column) BaBar data on the $\pi^0$ form-factor.
    We fit the parameters $M_V$, $e_m^V$ and the mixing couplings together with $\mP_1$ and $\mP_2$, not $\bar{d}_2$ and $\bar{d}_{123}$; these are reconstructed from the fitted values for $\mP_1$ and $\mP_2$ and eqs.~(\ref{eq:uncorrelate1}) and~(\ref{eq:uncorrelate2}) and shown here for illustration.
    The third column shows the parameters fitted excluding BaBar $\pi^0$ data and fixing $M_V$ and $e_m^V$ (marked with $\dagger$) according to refs.~\cite{MassSplitting}.
    The last row collects the $\chi^2$ per degree of freedom (dof).
    }}
    \label{tab:Fitted param}
\end{table}

 \begin{table}[!t]
     \centering\begin{tabular}{ccccccccc}\hline
     &$\mathcal{P}_1$&$\mathcal{P}_2$&$M_V$&$e_m^V$&$\theta_8$&$\theta_0$&$f_8$&$f_0$\\\hline\hline
    $\mathcal{P}_1$&1 & 0.085 & 0.511 & 0.638 & 0.495 & 0.017 & -0.107 & -0.025\\
    $\mathcal{P}_2$&0.085 & 1 & 0.434 & 0.439 & -0.157 & -0.616 & -0.058 & 0.434\\
    $M_V$&0.511 & 0.434 & 1 & 0.444 & 0.321 & -0.054 & -0.129 & 0.351\\
    $e_m^V$&0.638 & 0.439 & 0.444 & 1 & -0.081 & 0.059 & -0.179 & 0.319\\
    $\theta_8$&0.495 & -0.157 & 0.321 & -0.081 & 1 & -0.046 & -0.486 & -0.184\\
    $\theta_0$&0.017 & -0.616 & -0.054 & 0.059 & -0.046 & 1 & -0.020 & -0.424\\
    $f_8$&-0.107 & -0.058 & -0.129 & -0.179 & -0.486 & -0.020 & 1 & -0.156\\
    $f_0$&-0.025 & 0.434 & 0.351 & 0.319 & -0.184 & -0.424 & -0.156 & 1\\
\hline\hline
     \end{tabular}
     \caption{{\small Correlation matrix for our best fit (excluding BaBar $\pi^0$--TFF data).}}
     \label{tab:CorrWoB}
     \end{table}

We now focus on the fit to TFF data and eqs.~(\ref{eq:mix-input1})--(\ref{eq:mix-input4}).
A preliminary fit with $f_{8/0}$, $\theta_{8/0}$, $M_V$, $e_m^V$ and
$\bar{d}_2$ and $\bar{d}_{123}$
shows a correlation between
the latter two close to one.
Hence, instead of considering $\bar{d}_2$ and $\bar{d}_{123}$,
we will fit the parameters $\mP_1$ and $\mP_2$ provided by the linear combinations
\begin{subequations}
\begin{align}
    \bar{d}_2
    =&\alpha_2+\frac{\sigma_{d_2}}{\sqrt{2}}\left(\sqrt{1+r}\hspace*{1ex}\mathcal{P}_1 - \sqrt{1-r}\hspace*{1ex}\mathcal{P}_2\right),
\label{eq:uncorrelate1}
\\
    \bar{d}_{123}
    =&\alpha_{123}+\frac{\sigma_{d_{123}}}{\sqrt{2}}\left(\sqrt{1+r}\hspace*{1ex}\mathcal{P}_1 + \sqrt{1-r}\hspace*{1ex}\mathcal{P}_2\right),
\label{eq:uncorrelate2}
\end{align}
\end{subequations}
 with $\alpha_2 = -0.0272$, $\alpha_{123}=-0.233$,  $\sigma_{d_2}=10.49\cdot10^{-3}$, $\sigma_{d_{123}}=81.28\cdot10^{-3}$ and $r=0.995$.
 A fair estimate of the constants $\sigma_{d_2}$, $\sigma_{d_{123}}$ and $r$
 that minimize
 these fit correlations can be extracted from the errors and correlation between $\bar{d}_2$ and $\bar{d}_{123}$,
 respectively, in the preliminary fit.
 This transformation greatly reduces
 the correlation between both fit parameters, improving the efficiency and reliability of the numerical fit.
 We note that these constants $\sigma_{d_2}$, $\sigma_{d_{123}}$ and $r$ will change if one considers a different data set or fixes some parameters, e.g., $M_V$ and $e_m^V$.
 In this case, the fit to all the available space-like TFF data casts
 the mean values and errors given in table~\ref{tab:Fitted param} (first column of results;
 $\bar{d}_{2,123}$ were reconstructed from $\mP_{1,2}$ by means of eqs.~(\ref{eq:uncorrelate1}) and~(\ref{eq:uncorrelate2})).
 Further details on the correlations for this `fit~1'have been relegated to table~\ref{tab:CorrWB} in App.~\ref{app:correlations}. \\

However, this first fit points out that the $\pi^0$ BaBar data~\cite{BaBar} is in conflict with $\eta$ and $\eta'$-TFF data, related through chiral symmetry, and with Belle data (see e.g.
refs.~\cite{Mikhailov:2009kf, Roberts:2010rn, Brodsky:2011yv, Bakulev:2011rp, Brodsky:2011xx, Stefanis:2012yw, Bakulev:2012nh, Raya:2015gva, Eichmann:2017wil}).
Likewise, it is at odds with the asymptotic $\pi^0$ form-factor behaviour $\mF_{\pi^0\gamma^\star\gamma^\star}\sim \cO(Q^{-2})$ expected in high-energy QCD~\cite{BL}.
Indeed, this set of $\pi^0$ data is exclusively responsible for the important deviation in $\chi^2$ ($150.$) from the number of degrees of freedom of this `fit~1' ($101$).
Once the BaBar $\pi^0$ data set is removed from the analysis, the ratio $\chi^2/$dof drops
and our theoretical expressions provide a good description for all the $\pi^0$, $\eta$ and $\eta'$ experimental data.
Thus, we will take this `fit~2' as our best fit. The corresponding mean values and marginal standard deviations are given in table~\ref{tab:Fitted param} (second column of results) and their correlations can be found in table~\ref{tab:CorrWoB}.
In order to reduce the correlations in`fit~2' ($\pi^0$ BaBar data excluded), we have used transformations~(\ref{eq:uncorrelate1}) and~(\ref{eq:uncorrelate2}) with the same values for $\sigma_{d_2}$, $\sigma_{d_{123}}$ and $r$ as `fit~1' ($\pi^0$ BaBar data included). \\

\begin{figure}[!t]
\begin{center}
\includegraphics[width = .7\textwidth]{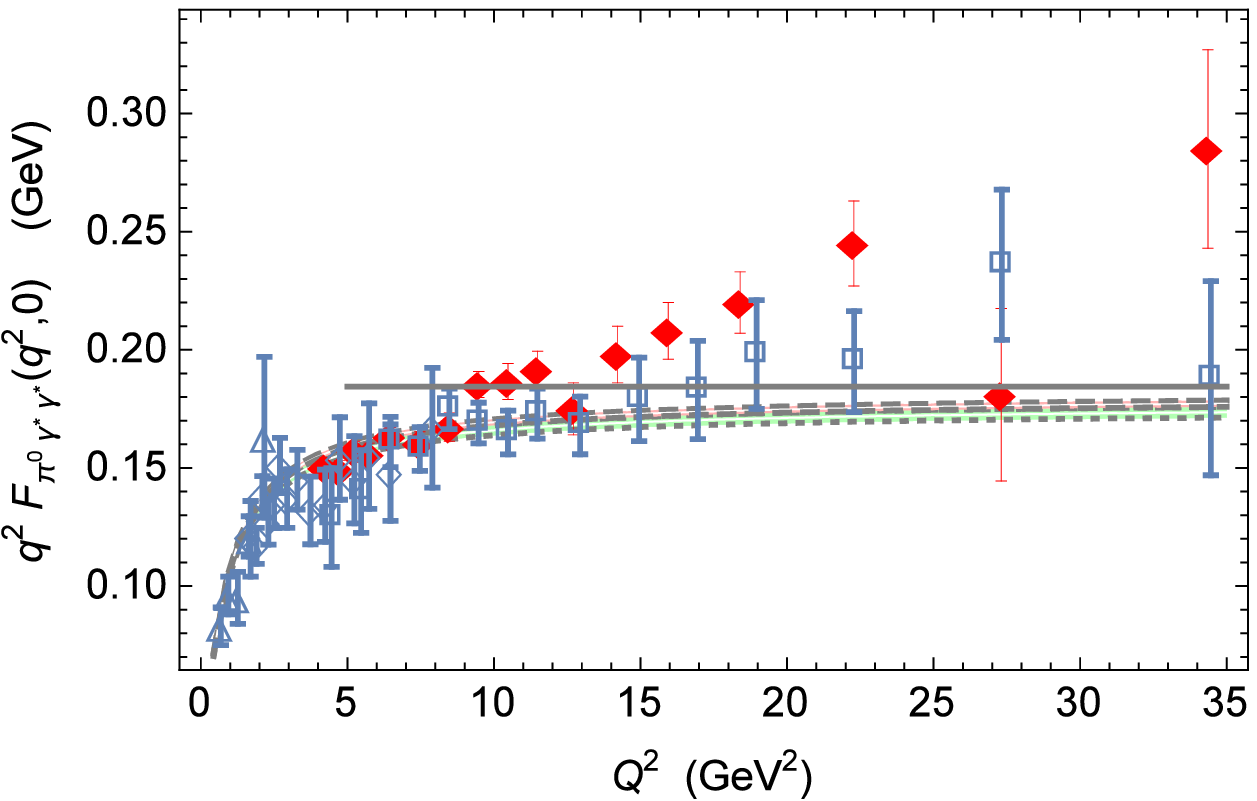}\\\vspace*{0.05cm}
\includegraphics[width = .7\textwidth]{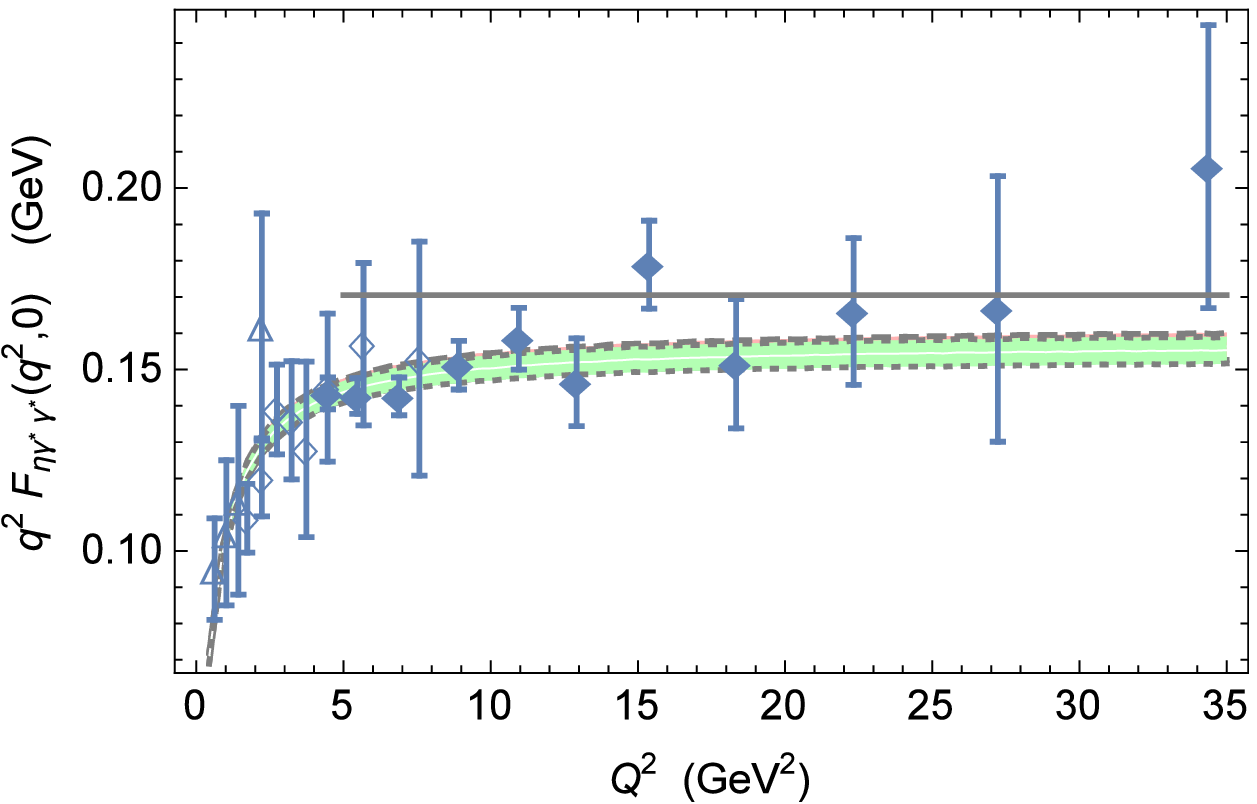}\\\vspace*{0.05cm}
\includegraphics[width = .7\textwidth]{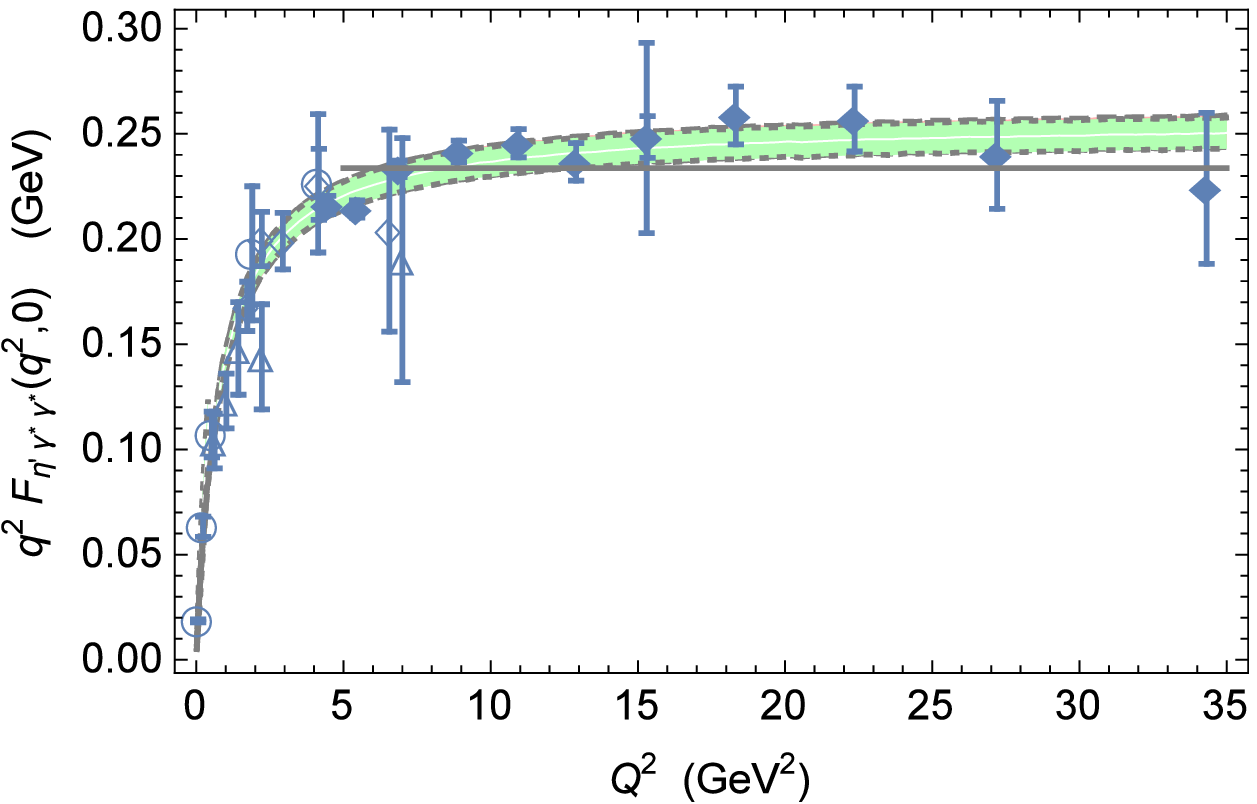}
\end{center}
\caption{{\small Comparison between the $\pi^0$ (top), $\eta$ (middle) and $\eta'$ (bottom) TFFs from the fit to all data (darker red band with dashed borders) and the fit after removing BaBar $\pi^0$ data (clearer green band with dotted borders). See the text for details.
    }}
\label{fig:noBaBar-vs-all}
\end{figure}
\begin{figure}[!t]
\begin{center}
\includegraphics[width = .695\textwidth]{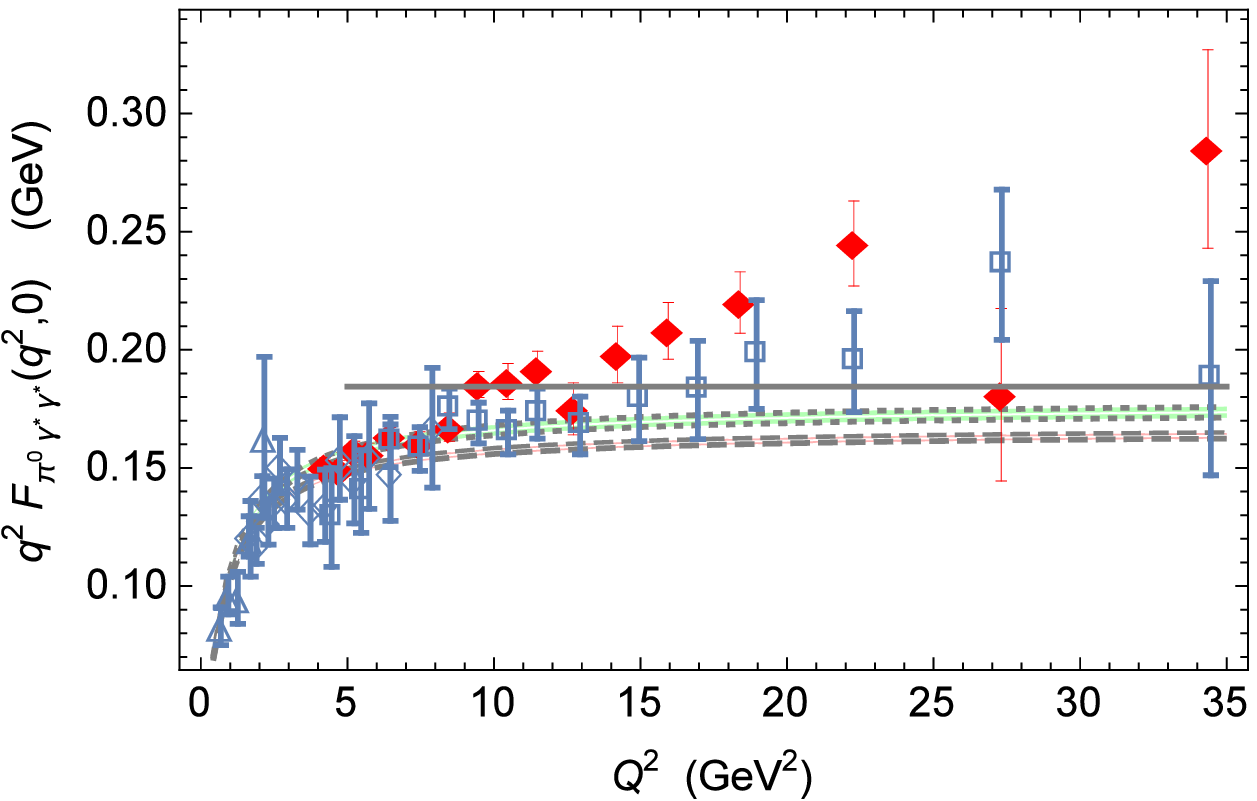}\\
\includegraphics[width = .695\textwidth]{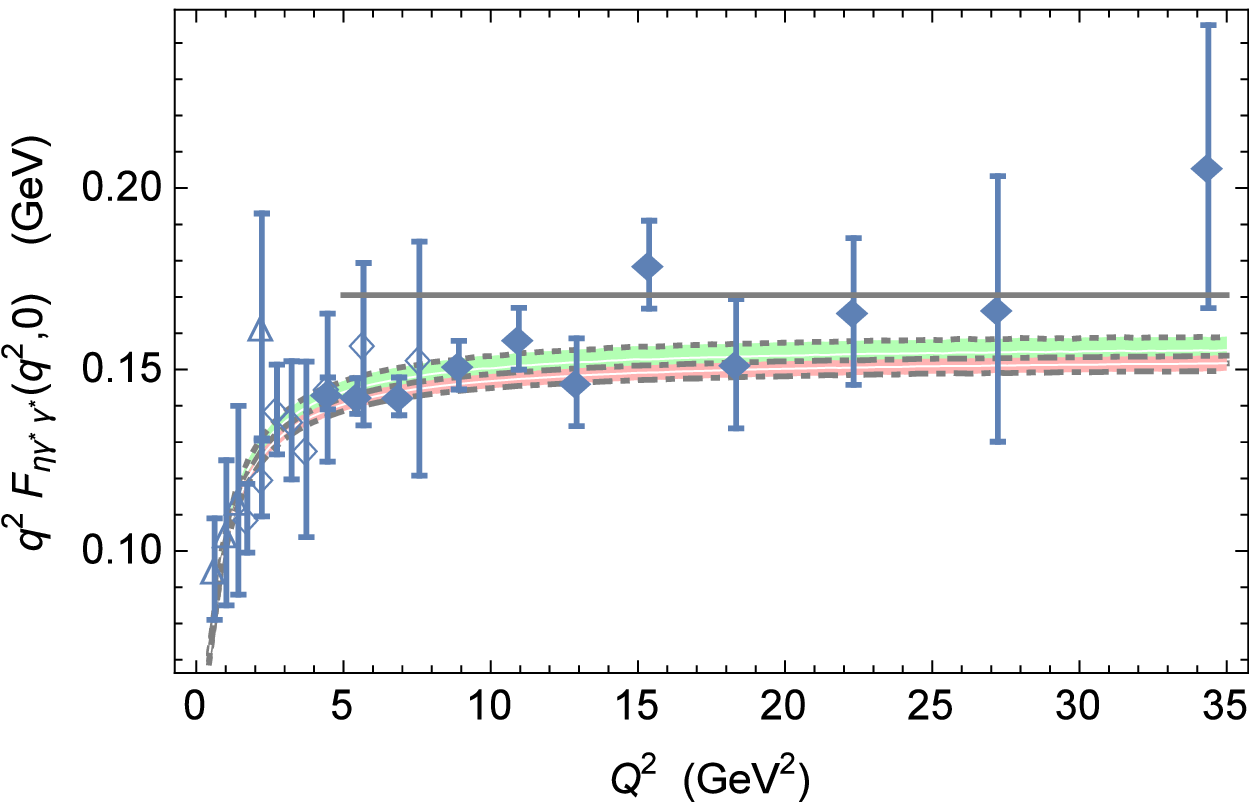}\\
\includegraphics[width = .695\textwidth]{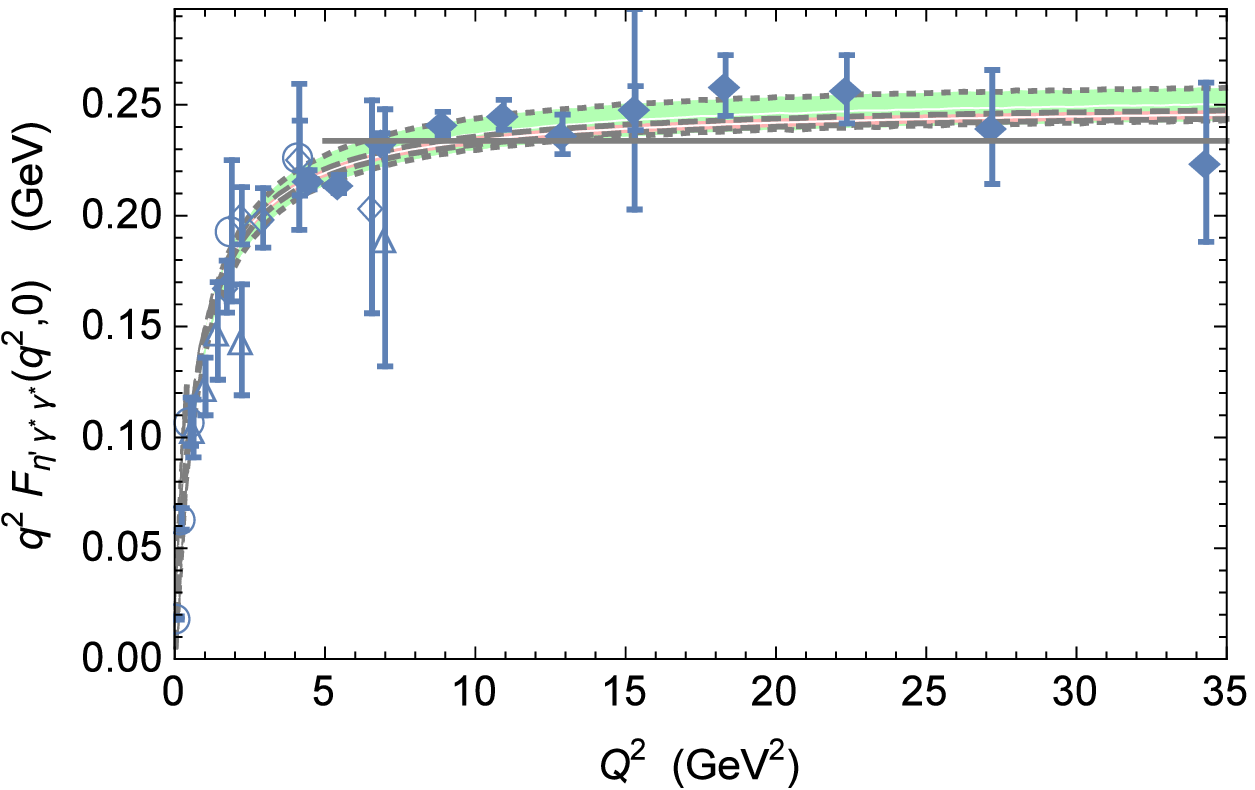}
\end{center}
\caption{{\small Comparison between the  $\pi^0$ (top), $\eta$ (middle) and $\eta'$ (bottom) TFFs from the fit with $M_V$ and $e_m^V$ fixed (darker red band with dashed borders) and
with these parameters included in the fit (clearer green band with dotted borders). BaBar $\pi^0$ data is not fitted in both cases. The remaining notation is the same as in fig.~\ref{fig:noBaBar-vs-all}.}}
\label{fig:noBaBar-vs-MV}
\end{figure}

 The values of $M_V$ obtained from these fits seem rather at odds with what one expects from the measured mass of the lowest-lying resonances: $M_\rho=M_\omega\approx 807$~MeV and $M_\phi\approx 1126$~MeV.
 In principle, this is not a problem for our analysis, as we are considering a purely space-like analysis within the lightest resonance multiplet and large--$N_C$ approximation. Thus, the parameters employed here are not truly
 those of the full theory with the large--$N_C$ infinite tower of resonances.
 Thus, the lightest vector couplings become slightly shifted to compensate for the heavier resonances missing in our R$\chi$T description.
 This could mean that resonances from heavier multiplets ({\it e.g.}, $\rho'$) might be giving a contribution to the experimental data of the form-factor which is not fully negligible in our approach. Therefore, we decided to make yet another fit fixing the mass and the $U(3)$ splitting parameter, excluding the $\pi^0$ data from BaBar, in order to study how this affects our results. This `fit~3' gives the values shown in the third column of table~\ref{tab:Fitted param}, and the correlation among parameters in table~\ref{tab:CorrMV} in app.~\ref{app:correlations}. As done in the previous fits, the
 $\bar{d}_2$ and $\bar{d}_{123}$
 parameters were transformed in the same way, however, for this fit the values of the decorrelating transformation constants read $\alpha_2=-2.95\cdot10^{-2}$, $\alpha_{123}=-0.254$, $\sigma_{d_2}=6.88\cdot10^{-3}$, $\sigma_{d_{123}}=5.04\cdot10^{-2}$ and $r = 0.992$. The comparison of the second and third columns of results in table~\ref{tab:Fitted param} shows only small changes in the fitted parameters and the fit quality, which supports the values obtained for $M_V$ and $e^V_m$ in the first two fits of table~\ref{tab:Fitted param}. Based on this discussion, we consider the second column of results in table~\ref{tab:Fitted param} as our reference fit.\\

In fig.~\ref{fig:noBaBar-vs-all}, we show the comparison between our best fit (excluding the BaBar $\pi^0$ TFF) and the fit to all the data ($Q^2=-q^2$).
The $\pi^0$-TFF from the fit to all data is represented by the darker red band with dashed borders and the fit after removing BaBar $\pi^0$ data is given by the clearer green band with dotted borders. These bands provide the TFF  $1\sigma$ uncertainty stemming from the corresponding fit, taking into account correlations. These results are shown together with the experimental data from BaBar~\cite{BaBar,BABAR:2011ad} (full diamonds), Belle~\cite{Belle} (squares), CLEO~\cite{CLEO} (empty diamonds) and CELLO~\cite{CELLO} (triangles).
The outcomes from both fits are very similar for the $\pi^0$ TFF and essentially identical for the $\eta$ and $\eta'$ form-factors.
We have also plotted the QCD asymptotic $1/q^2$ coefficient $\lim_{Q^2\to\infty} q^2 \mF_{P\gamma^\star\gamma^\star}(q^2,0)$ in fig.~\ref{fig:noBaBar-vs-all} (horizontal line),
estimated by $2 F \times C_{P_0}$ with the central values from our best fit and $F\approx F_\pi$ for illustration~\cite{BL} \footnote{A detailed analysis of the singlet contribution \cite{Agaev:2014wna} tends to reduce (increase) slightly the value for the asymptotic $1/q^2$ coefficient in the $\eta$ ($\eta'$) TFFs (see also the discussion in Ref.\cite{Escribano:2015yup}), in closer agreement with the trend shown by data.}.
Since the experimental TFFs (barring BaBar $\pi^0$ data) are compatible with this asymptotic behaviour, our theoretical form-factors also approximately agree with it,
even though the matching of the asymptotic $1/q^2$ coefficient from QCD
was not among our high-energy constraints.\\

The comparison of our best fit with that disregarding the $\pi^0$ data from BaBar but fixing $M_V$ and $e_m^V$ is given in fig.~\ref{fig:noBaBar-vs-MV}.
The TFF from the fit with $M_V$ and $e_m^V$ fixed is given by the darker red band with dashed borders. It is compared to our best fit, where $M_V$ and $e_m^V$ are fitted
together with the other six parameters, provided by the clearer green band with dotted borders.
The remaining notation is the same as in the previous plots in fig.~\ref{fig:noBaBar-vs-all}.
The results for the $\eta$ and $\eta'$ TFFs happen to be compatible --though less precise for our best fit--. On the other hand, there is some small discrepancy in the $\pi^0$-TFF:
the $1\sigma$ bands do not overlap and the asymptotic value for the $\pi^0$ form-factor is slightly smaller in the case with fixed $M_V$ and $e_m^V$, yielding a poorer $\chi^2/$dof.
Notice that the worsening of the fixed--$(M_V,e_m^V)$ fit is due to the $\pi^0$--TFF.\\

 The departure of all three values for $\bar{d}_{123}$
 in table~\ref{tab:Fitted param} from the prediction $d_{123}=1/24\approx 4\cdot10^{-2}$ \cite{SDConst} shows the impact
 of pseudoscalar resonances in this particular coupling. Our fitted values of
 $\bar{d}_2$
 exhibit a similar deviation with respect to those obtained in ref.~\cite{Dai:2013joa}, where $d_2\in[4,7]\cdot10^{-2}$ (see also ref.~\cite{Chen:2012vw}),  pointing again to the relevance of pseudoscalar resonance contributions to $VVP$ vertices.\\

   We also evaluated the mixing parameters $C_q^{(\prime)}$ and $C_s^{(\prime)}$ according to our best fit (BaBar $\pi^0$ data excluded).
   Since our fit determines $f_8$ and $f_0$ rather than their respective ratios $f_8/F$ and $f_0/F$ with the chiral limit decay constant $F$, we define the quantities
   $\overline{C}_{q/s}^{(')}\equiv C_{q/s}^{(')} \times (F_\pi/F)$.
   These quantities only depend now on the angles $\theta_{8/0}$ and the ratios $f_{8/0}/F_\pi$ and can be easily translated into the actual mixing coefficients $C_{q/s}^{(')}$ once the value of the ratio $F_\pi/F$ is provided.
   They are given in table~\ref{tab:mixing}, where the mean values, the marginal standard deviations and the correlation matrix are quoted. The errors have been propagated from the fit parameters taking into account correlations by means of a MonteCarlo.\\

   \begin{table}[!t]
       \centering
       \begin{tabular}{c|cccc|cc}\hline
              & $\overline{C}_q$ & $\overline{C}_s$ & $\overline{C}_q'$ & $\overline{C}_s'$ & Mean value $\pm1\sigma$&Czy\.z {\it et al.} \cite{Czyz}\\\hline\hline &&&&& \vspace*{-0.25cm}\\
         $\overline{C}_q$  & 1 & 0.404 	
         & 0.334
         & -0.469
         & $0.69\pm0.03$
         & $0.61\pm0.05$
         \\
         $\overline{C}_s$ & 0.404
         & 1 & 0.127	
         & 0.008
         & $0.47\pm0.03$
         & $0.14\pm0.06$
         \\
         $\overline{C}_q'$ & 0.334
         & 0.127
         & 1 & -0.820
         & $0.60\pm0.03$
         & $0.56\pm0.06$
         \\
         $\overline{C}_s'$ & -0.469
         & 0.00754
         & -0.820
         & 1
         & $0.58\pm0.02$
         & $0.74\pm0.08$
         \\\hline\hline
       \end{tabular}
       \caption{Correlation matrix for the $\eta-\eta'$ mixing parameters, mean values and marginal standard deviations according to our best fit analysis.
       These results are compared to the corresponding values from ref.~\cite{Czyz}.}
       \label{tab:mixing}
   \end{table}

\subsection{Comparison with other recent TFF determinations} \label{Comparisons}

   Our description of the Transition Form Factor provides fairly simple expressions that can be implemented in the computation of $P\gamma\gamma$ observables, despite the lack of NLO contributions in the $1/N_C$ expansion (barring those in the $\eta-\eta'$ mixing). This derives from the also simple short distance constraints~(\ref{eq:new-SD-rels}), contrary to those in ref.~\cite{Czyz}, where much more involved ultraviolet restrictions are reported. These are the result of a more general $U(3)$ breaking pattern, not introduced in the chirally covariant form proposed in this article, and the inclusion of several vector multiplets. Their fits are done to observables in both $q^2$ regimes, time-like and space-like, without considering photon radiative corrections~\cite{NLODD} which may become important for some observables.\\

   \begin{table}[!ht]
       \centering
       \begin{tabular}{cc}\hline
       Czy\.z {\it et al.} \cite{Czyz}& This work\\\hline\hline
       $f_{V_1} $ &$ \frac{1}{M_{\rho}}(F_V+8m_\pi^2\lambda_V) $
       \\
       $F_{\omega_1}$
       & 1
       \\
       $F_{\phi_1}$
       &   $ \frac{M_\rho}{M_{\phi}} \frac{ (F_V+8\Delta_{2K\pi}^2\lambda_V)}{(F_V+8m_\pi^2\lambda_V)} $
       \\
       $h_{V_1}^{\pi_0} $&$\frac{1}{ M_VM_{\rho}}(-c_{1256}M_\rho^2+  c_{1235}^\star  m_{\pi}^2)$\\
       $h_{V_1}^{\eta} $&$\frac{1}{ M_VM_{\rho}}(-c_{1256}M_\rho^2+   c_{1235}^\star  m_{\eta}^2-8  c_3^\star   \Delta_{\eta\pi}^2)$\\
       $H_{\omega_1}$&1\\
       $A^{\pi_0}_1$&0\\
       ${h_{V_1}^\eta } \left[2C_s-\left(\frac{5}{\sqrt{2}}C_q-C_s\right)A^{\eta}_1\right] $
       &$ \frac{C_s}{ M_VM_{\phi}}(-c_{1256}M_{\phi}^2 +  c_{1235}^\star  m_{\eta}^2-8 c_3^\star   \Delta_{2K\pi\eta}^2)$\\
       $\sigma_{V_1}^{\pi_0} $ &$-\frac{1}{M_{\rho} M_{\omega}}\left[d_{123}^\star  m_{\pi}^2+  d_3^\star   (M_\rho^2+M_\omega^2)\right]$\\
       $\sigma_{V_1}^{\eta} $ &$-\frac{1}{M_{\rho}^2}\left[   d_{123}^\star   m_{\eta}^2-  d_2^\star  \Delta_{\eta\pi}^2+2 d_3^\star  M_\rho^2\right]$\\
       $A^{\pi^0}_{\phi\omega,1}$&0\\
       $A^{\eta}_{\phi\omega,1}$&0\\
       $\frac{\sigma_{V_1}^\eta}{ F{_{\phi_1}}}\left[5C_qA_1^\eta-\sqrt{2}C_s(A_1^\eta+2)\right]$
       &$\frac{2C_s}{M_{\phi}^2}\left[ d_{123}^\star  m_\eta^2- d_2^\star \Delta_{2K\pi\eta}^2+2 d_3^\star  M_\phi^2\right]$\\\hline\hline
       \end{tabular}
       \caption{Relations between the parameters in the model of ref.~\cite{Czyz} and this work. The expressions have been derived comparing various on-shell vertices, as explained in the text. The relations for $\eta'$ are again obtained from the $\eta$ ones as explained after eq.~(\ref{etaTFF}).}
       \label{tab:Czyz}
   \end{table}

   This model is analogous to ours in the sense that all the parameters in our model can be expressed in terms of those in ref.~\cite{Czyz},  when restricted to the first multiplet given by $i=1$ \footnote{In this regard, it is interesting to note that the earlier analysis by Czyz \textit{et. al.} \cite{Czyz:2012nq} did not need to include these higher radial excitations when fitting only space-like data.}. In our approach the Lagrangian couplings are quark mass independent and the structure of $m_{q/s}$ corrections is dictated by the operators and the kinematics ($(q_1+q_2)^2=p^2=m_P^2$). On the other hand, the effective parameters in~\cite{Czyz} encoded an important part of the quark mass corrections.
   Note, however, that the $U(3)$ splitting in ref.~\cite{Czyz} is not fully general and assumes some restrictions: their vector-photon-pseudoscalar ($h_{V_i}$) and vector-vector-pseudoscalar ($\sigma_{V_i}$) couplings are the same for the three pseudo-Goldstones. The equivalences are given in table~\ref{tab:Czyz}.
   In order to do the comparison, we have translated our resonance Lagrangian in the antisymmetric tensor $V_{\alpha\beta}$ formalism into the Proca four-vector $\hat{V}_\rho$ realization considered in~\cite{Czyz} through
   the transformation $V_i^{\mu\nu}\to -(\partial^\mu \hat{V}_i^\nu -\partial^\nu \hat{V}_i^\mu)/M_{V_i}$ for each vector $V_i=\rho,\omega,\phi,\rho'...$~\cite{Bijnens:1995ii,Kampf:2006yf}. Once our $\mL_{R\chi T}$ is expressed in terms of Proca fields, the operators are set with their particles on-shell and identified with the corresponding notation in~\cite{Czyz}.\\

    In table~\ref{tab:Czyznum} we give the numerical values of the parameters in table~\ref{tab:Czyz} for both descriptions ('fit 2' results are used in the comparison). It can be seen that,  in spite of the differences in both descriptions, a very good agreement is found among the numerical values for table~\ref{tab:Czyznum}. It is worth mentioning that, despite the very different results for $\theta_8$, $\theta_0$, $f_8$ and $f_0$, the mixing parameters $C_q^{(\prime)}$ are in good agreement but $C_s$ and $C'_s$ show some discrepancy (more significant for the latter), as it can be seen in table~\ref{tab:mixing}. However, if the relations of eqs. (\ref{stabilizer}) are not imposed in our fits, the values obtained for the mixing parameters in this way get closer to those in ref.~\cite{Czyz}.\\

   \begin{table}[!ht]
       \centering
       \begin{tabular}{ccc}\hline
       &Czy\.z {\it et al.} \cite{Czyz}& This work\\\hline\hline
       $f_{V_1} $ &  $0.2020\pm0.0008$ & $0.198\pm0.003$
       \\
       $F_{\omega_1}$
       & $0.88\pm0.01$& 1
       \\
       $F_{\phi_1}$
       &   $0.783\pm0.005$ & $0.72\pm0.04$
       \\
       $h_{V_1} $& $0.0377\pm0.0008$ & $0.0326\pm0.0005$
       \\
       $H_{\omega_1}$&$1.02\pm0.03$&1\\
       $A^{\pi_0}_1$&$-0.083\pm0.002$&0\\
       ${h_{V_1}^\eta } \left[2C_s-\left(\frac{5}{\sqrt{2}}C_q-C_s\right)A^{\eta}_1\right] $
       &$0.39\pm0.09$&$0.36\pm0.07$\\
       $\sigma_{V_1}^{\pi} $ &$0.264\pm0.007$&$0.240\pm0.008$\\
       $\sigma_{V_1}^{\eta} $ &$0.264\pm0.007$&$0.2\pm0.3$\\
              $\sigma_{V_1}^{\eta'} $ &$0.264\pm0.007$&$0.3\pm0.8$\\
       $-{h_{V_1}^{\eta'} } \left[2C_s'+\left(\frac{5}{\sqrt{2}}C_q'+C_s'\right)A^{\eta'}_1\right] $
       &$-0.37\pm0.11$&$-0.45\pm0.07$\\
       $A^{\pi^0}_{\phi\omega,1}$&$-0.21\pm0.04$&0\\
       $A^{\eta}_{\phi\omega,1}$&$-0.027\pm0.007$&0\\
       $\frac{\sigma_{V_1}^{\eta}}{ F{_{\phi_1}}}\left[5C_qA_1^\eta-\sqrt{2}C_s(A_1^\eta+2)\right]$
       &$-0.42\pm0.03$&$-0.27\pm0.02$\\
       $\frac{\sigma_{V_1}^{\eta'}}{ F{_{\phi_1}}}\left[5C_q'A_1^{\eta'}+\sqrt{2}C_s'(A_1^{\eta'}+2)\right]$
       &$0.43\pm0.16$&$0.33\pm0.05$\\\hline\hline
       \end{tabular}
       \caption{Numerical values for the linear combinations of constants shown in table~\ref{tab:Czyz},  reported in ref.~\cite{Czyz},  compared to the values obtained from our analysis.}
       \label{tab:Czyznum}
   \end{table}
%

   On the other hand, descriptions of the TFF such as those in ref.~\cite{Masjuan:2017tvw} by means of Pad\'e approximants provide a neat and simple approach, which
   can also incorporate asymptotic QCD information~\cite{BL,Nesterenko}.
   However, the lack of a Lagrangian in such method makes it complicated to combine information from the $\pi^0$ and $\eta^{(')}$ TFFs.
   As a result, one needs to perform a separate Pad\'e analysis for each quantity, needing a set of similar statistical quality data to describe a closely related process with a comparable accuracy.

\subsection{$\mB(P\to\gamma^{(\star)}\gamma^{(\star)})$ predictions vs. data}\label{subsec:BR}
\label{Predictions}

  As a quality check of the fitted parameters and the R$\chi$T description, we give our estimates for different branching ratios related 
  to $P\to\gamma^{(\star)}\gamma^{(\star)}$ processes, uncertainties 
  {are naively obtained by independently varying the R$\chi$T parameters within their $1\sigma$ ranges 
  without considering correlations among them}~\footnote{
  {We will study these and other related processes in a forthcoming paper \cite{NextPaper}, 
  where uncertainties and their correlations will be discussed in detail.   }  
  }. These are provided by our best fit in the table~\ref{tab:DecWidth}
  (all parameters are floated and BaBar $\pi^0$ data is excluded).
  The width of the resonances is needed to obtain finite results. A momentum dependent width is employed for the $\rho$ meson, following ref.~\cite{GomezDumm:2000fz}; for the $\omega$ and $\phi$, we use the total widths from PDG~\cite{PDG} neglecting any off-shellness dependence, as they are rather narrow. This is accomplished by changing the denominator of the corresponding resonance propagator to $D_R(s)=(M_R^{\rm phys})^ 2-s-i M_R\Gamma_R(s)$, with the physical masses $M_R^{{\rm phys}}$ taken from the PDG~\cite{PDG}. Except for the resonances masses in the propagators $D_R(s)$, all parameters correspond to those from our best fit.
  For the computation of our predicted branching ratios we have divided our R$\chi$T results for the partial widths by the experimental total decay width (no theoretical prediction is considered here for the latter).\\

  \begin{table}[!ht]
      \centering
      \begin{tabular}{c c c}\hline
         Process & Predicted branching fraction & PDG value\\\hline\hline
    $\pi^0\to e^+e^-\gamma$ & $(1.16\pm0.06)\cdot10^{-2}$ & $1.174(35)\cdot10^{-2}$\\
    $\eta\to e^+e^-\gamma$ & $(7.1\pm0.8)\cdot10^{-3}$ & $6.9(4)\cdot10^{-3}$\\
    $\eta\to \mu^+\mu^-\gamma$ & $(3.4\pm0.3)\cdot10^{-4}$ & $3.1(4)\cdot10^{-4}$\\
    $\eta'\to e^+e^-\gamma$ & $(5.3\pm1.1)\cdot10^{-4}$ & $4.73(30)\cdot10^{-4}$\\
    $\eta'\to \mu^+\mu^-\gamma$ & $(1.3\pm0.3)\cdot10^{-4}$ & $1.09(27)\cdot10^{-4}$\\
    $\pi^0\to 2e^+2e^-$ & $(3.24\pm0.16)\cdot10^{-5}$ & $3.34(16)\cdot10^{-5}$\\
    $\eta\to 2e^+2e^-$ & $(2.4\pm0.3)\cdot10^{-5}$ & $2.40(22)\cdot10^{-5}$\\
    $\eta\to 2\mu^+2\mu^-$ & $(4.0\pm0.4)\cdot10^{-9}$ & $<3.6\cdot10^{-4}$\\
    $\eta\to e^+e^-\mu^+\mu^-$ & $(2.4\pm0.2)\cdot10^{-6}$ & $<1.6\cdot10^{-4}$\\
            $\eta'\to 2e^+2e^-$ & $(2.2\pm0.5)\cdot10^{-6}$ & No bounds\\
        $\eta'\to 2\mu^+2\mu^-$ & $(2.2\pm0.5)\cdot10^{-8}$ & No bounds\\
    $\eta'\to e^+e^-\mu^+\mu^-$ & $(1.2\pm0.2)\cdot10^{-7}$ & No bounds\\
    \hline\hline
      \end{tabular}
      \caption{Predicted branching fractions of the fitted parameters excluding BaBar $\pi^0$ data compared to PDG data. PDG upper bounds are given at the 90 $\%$ confidence level.}
      \label{tab:DecWidth}
  \end{table}

   The second and third column of table~\ref{tab:DecWidth} agree at less than two standard deviations in all cases (theory undertainties are not discussed in these estimates), which corroborates our best fit results. We note that no experimental limit is known for the decays
   in the last three lines of table~\ref{tab:DecWidth}.
   We emphasize that our space-like analysis is missing several features which may be crucial in time-like observables, such as subleading $1/N_C$ effects as, e.g., resonance widths and the impact of higher resonance multiplets. Thus, the R$\chi$T branching ratios in table~\ref{tab:DecWidth} should be taken with a grain of salt. Nonetheless, the fair agreement with data is remarkable, giving  support to the hypotheses and approximations assumed in this work and the final results for the anomalous magnetic moment in the next section. \\

\section{Pseudo-Goldstone pole contribution  to $a_\mu^{HLbL}$}\label{a_mu}
\label{sec:g-2}

{\subsection{Pole prediction with one vector resonance multiplet }
}

For the evaluation of the pseudoscalar pole contribution to the HLbL, $a_\mu^{P,HLbL}$, we used the expressions for the loop integrals given in ref.~\cite{Knecht}~\footnote{The specific formulae used can be found in appendix \ref{app:Formulae}.}. Dispersion relations~\cite{Bern,Mainz} show that the HLbL is determined by the various physical absorptive channels of the $VVVV$ Green's function of four electromagnetic currents. The lightest absortive cut is given by the meson pole topologies  $\gamma^\star\gamma^\star\to P\to \gamma^\star\gamma^\star$. Additional intermediate states ($PP$ cuts, multiparticle states or with heavy resonances, etc.) will add further corrections to $a_\mu^{HLbL}$.
These were effectively included before in the so-called meson-exchange contributions (with off-shell mesons) \cite{Jegerlehner}. In a dispersive framework, however, the analyticity structure of the amplitudes is encoded in their poles and cuts alone, in such a way that their residues and imaginary parts (discontinuities) are uniquely related to on-shell form
factors and scattering amplitudes~\cite{Bern, Mainz}.\\

Notice that the $\mathcal{O}(m_\pi^2)$ correction to the form-factor from considering a non-vanishing pseudo-Goldstone mass given in ref.~\cite{us} is automatically included in our analysis. Since no $U(3)$-splitting parameter was considered in the analysis of the form-factor in ref. \cite{us}, such correction was underestimated. This contribution accounts for a relative variation of the form-factor of $\Delta\sim-2.5\cdot10^{-2}$,~\footnote{The precise definition of $\Delta$ can be found in ref.~\cite{us}.}
while in ref. \cite{us} this correction is reported to be $\Delta\sim5.9\cdot10^{-3}$. However, as stated there, additional corrections come as further suppressed powers of $m_P^2$. \\

The total pseudo-Goldstone pole contribution is estimated by
means of a MonteCarlo run with 5000 events, which randomly generates  the eight fit parameters with a normal distribution according to their mean values, errors and correlations between parameters.
It integrates at the same time all three contributions from $\pi^0$, $\eta$ and $\eta'$ exchanges, thus taking into account the correlations between the $\pi^0$, $\eta$ and $\eta'$ TFF contributions.
Our best fit (`fit 2' --excluding BaBar $\pi^0$ data--) leads to our final result
 \begin{equation}\label{referenceresult}
     a_\mu^{P,HLbL} = (8.47\pm 0.16)\cdot10^{-10}\, ,
 \end{equation}
which perfectly agrees with previous results \cite{Kampf, us, Masjuan:2017tvw, Knecht, Hayakawa:1997rq, Bijnens:2001cq, Melnikov:2003xd, Erler:2006vu, Hong:2009zw, Cappiello:2010uy, Goecke:2010if, Dorokhov:2011zf,
Masjuan:2012wy, Masjuan:2012qn, Escribano:2013kba,
Blum:2016lnc, Blum:2017cer} \footnote{Comparisons should be made with evaluations of the pseudoscalar on-shell pole contributions.} and with a reduced uncertainty. In general, all evaluations are in agreement, as can be seen from table~\ref{tab:Comparison}. Noticeable exceptions are (we consider the last result from every group as the reference one) those using Melnikov-Vainshtein short-distance constraints \cite{Melnikov:2003xd}, obtaining $\sim13.5\cdot10^{-10}$ and the results obtained within the non-local chiral quark model by Dorokhov \textit{et al.}, $\sim5.85\cdot10^{-10}$ \cite{Dorokhov:2011zf}. Although the lattice result of ref.~\cite{Blum:2016lnc} $(5.35\pm1.35)\cdot10^{-10}$
(see also refs. \cite{Green:2015sra, Gerardin:2016cqj, Bijnens:2016hgx, Asmussen:2017bup}),
might seem at odds with other determinations, one still has to take into account 
potentially large finite-volume systematics, finite-lattice-spacing corrections 
and the limited statistics for the leading disconnected contribution, all of which 
are being refined and could bring in reasonable agreement this result with the remaining predictions. From these numbers and comparing ours with those previously quoted, it seems that further orders in $m_P^2$ might be neglected in our analyses.

For comparison we also provide the pole contribution for the anomalous magnetic moment stemming from the other two types of fits done in this article:
 \begin{eqnarray}
 \mbox{Including BaBar $\pi^0$ data `fit 1':} \quad &&
 a_\mu^{P,HLbL} = (8.58\pm 0.16)\cdot10^{-10}.\;\;
\nn\\
\mbox{Fixing $M_V$ and $e_m^V$ `fit 3':}\quad&&a_\mu^{P,HLbL} = (8.50\pm 0.13)\cdot10^{-10},\;\;
 \end{eqnarray}
The values given by BaBar for the $\pi^0$-TFF at high energies are much larger than the form-factor derived from the $\eta$ and $\eta'$ TFFs through chiral symmetry. This leads to a slightly higher value for $a_\mu^{P,HLbL}$ in comparison to our best fit, with BaBar $\pi^0$ data excluded, which shows a very good compatibility between the $\pi^0$, $\eta$ and $\eta'$ data. Both results are still compatible within errors. On the other hand, the fit where the vector masses are fixed gives a $1\sigma$ confidence interval fully included in the $1\sigma$ interval of our best determination~(\ref{referenceresult}), although with a slightly smaller error, as expected.
 \begin{table}[!t]
     \centering\begin{tabular}{ll}\hline
     Reference &
     $10^{10}\, \cdot \,  a_\mu^{P,HLbL}$
     \\\hline\hline
     Knecht and Nyffeler \ (2002)  \cite{Knecht} &\hspace*{0.2cm} 8.3 \ \ $\pm$ \ 1.2\\
     Hayakawa and Kinoshita \ (2002)  \cite{Hayakawa:1997rq} &\hspace*{0.2cm}  8.3 \ \ $\pm$ \ 0.6 \\
     Bijnens, Pallante and Prades \ (2002) \cite{Bijnens:2001cq} &\hspace*{0.2cm}  8.5 \ \ $\pm$ \ 1.3\\
     Goecke, Fischer and Williams \ (2012) \cite{Goecke:2010if} &\hspace*{0.2cm}  8.1  \ \ $\pm$ \ 1.2
     \\
     Roig, Guevara and L\'opez Castro \ (2014) \cite{us} &\hspace*{0.2cm}  8.60 \ $\pm$ \ 0.25\\
     Masjuan and S\'anchez-Puertas \ (2017) \cite{Masjuan:2017tvw} &\hspace*{0.2cm}  9.4 \ \ $\pm$ \ 0.5\\
     Czy\.z, Kisza and Tracz \ (2018) \cite{Czyz} &\hspace*{0.2cm}  8.28\  $\pm$ \ 0.34\\
     \hline
     This work
     &\hspace*{0.2cm}  8.47\   $\pm$ \ 0.16 \\
\hline\hline
     \end{tabular}\caption{{\small Comparison of different predictions for the pseudoscalar-pole contributions to $a_\mu^{HLbL}$.
     }}
     \label{tab:Comparison}
     \end{table}
\vspace{0.1cm}
Given the abovementioned tension between QCD-driven predictions and BaBar $\pi^0$ transition form-factor data, we consider the result~(\ref{referenceresult}) excluding these data from the fits as our reference value, $a_\mu^{P,HLbL} = (8.47\pm 0.16)\cdot10^{-10}$.\\

From the fit considering the conditions given in eqs. (\ref{stabilizer}), but taking $F\approx87$ MeV (chiral limit) instead of $F_\pi$, we get $a_\mu^{P,HLbL}=(8.47\pm0.17)\cdot10^{-10}$. If, instead of the values of eqs. (\ref{stabilizer}), we use the $\eta-\eta'$ mixing conditions from the NNLO $U(3)$ $\chi$PT fit to lattice data~\cite{Guo:2015xva}, we get $a_\mu^{P,HLbL}=(8.57\pm0.16)\cdot10^{-10}$. These results are compatible with our reference value in eq. (\ref{referenceresult}), showing that our result is not fixed by the precise input given for the numerical values used to stabilize the  $\eta-\eta'$ mixing parameters. For the fit without the stabilizing conditions, eqs. (\ref{stabilizer}), we get $a_\mu^{P,HLbL}=(8.87\pm0.16)\cdot10^{-10}$, which is $2.5 \sigma$ away from our reference determination in eq. (\ref{referenceresult}),  $a_\mu^{P,HLbL}=(8.47\pm0.16)\cdot10^{-10}$. However, this result is not reliable since part of the phase space generated by a Gaussian random distribution of the parameters according to such fit are not physical and must be discarded~\footnote{Since a Gaussian distribution accounting for correlation is used to compute $a_\mu^{P,HLbL}$, there are some points of the generated phase space that give negative squared resonance masses, leading to spurious divergences. Therefore, these points were dropped in order to obtain a finite result.}. This shows the need to employ our stabilizing conditions for the $\eta-\eta'$ mixing in eqs. (\ref{stabilizer}).\\

Since the separate contributions of the $P$ mesons are interesting in their own ({\it i.e.}, some papers only consider the $\pi^0$ contribution, which is highly restricted by chiral symmetry, contrary to the $\eta^{(\prime)}$ cases; data quality varies from channel to channel, etc.), we quote their values in the following.
The pole contributions from each separate pseudo-Goldstone exchange are computed for our best fit via a 5000 event MonteCarlo in the same way described before,
giving the following values for our best fit:
\begin{subequations}
\begin{align}\label{individualcontributions}
    a_\mu^{\pi^0,HLbL} & = (5.81\pm0.09)\cdot10^{-10},\\
    a_\mu^{\eta,HLbL} &= (1.51\pm0.06)\cdot10^{-10},\\
    a_\mu^{\eta',HLbL} &= (1.15\pm0.07)\cdot10^{-10}\, .
\end{align}
\end{subequations}
Notice that despite the slightly higher (absolute) uncertainty in the $\pi^0$ contribution \footnote{Our
result for this contribution is in agreement at the 1.7 $\sigma$ level with the very recent dispersive evaluation of Hoferichter
\textit{et. al.} \cite{Hoferichter:2018dmo}, $a_\mu^{\pi^0,HLbL} = (6.26^{+0.30}_{-0.25})\cdot10^{-10}$.},
the $\eta$ has an uncertainty reduced by a factor of four compared with previous determinations \cite{us}. Also, the $\eta'$ contribution has a mildly reduced uncertainty. We note that the sum in quadrature (uncorrelated) of the errors for the individual contributions in the preceding equations yields an error estimate ($\sim0.13\cdot10^{-10}$), which is a bit smaller than the one in eq.~(\ref{referenceresult}) because of the correlations, accounted in the latter: performing the simultaneous integral of the three contributions leads to eq.~(\ref{referenceresult}).

As a check of the size of these corrections, we have additionally computed the $P$ pole contributions using  our form-factor in the chiral and large--$N_C$ limits. We have taken the central values of the vector mass $M_V$ and mixing parameters from our best fit, though keeping the physical pseudo-Goldstone masses in the integration kernels. As a result, we find $(F/F_\pi)^2\, \, a_\mu^{P,HLbL}=8.27\cdot10^{-10}$, where the change is essentially given by the $a_\mu^{\eta,HLbL}$ contribution. For $F\approx F_\pi$, the small change in the mean value of the pseudo-Goldstone pole contribution in the chiral and large--$N_C$ limits ($\sim - 2.5\%$, up to corrections in $F/F_\pi$) suggests that NNLO corrections, suppressed by further powers of $m_P^2$, must be negligible.
 An effect on this ballpark could already be inferred by comparing  $a_\mu^{P,HLbL}$ in the $U(3)$ symmetric analysis of ref.~\cite{us} with our present result. To check the compatibility between the present fit and that done in ref.~\cite{us} we compare the values obtained for
 the relevant combination $F_V^2 d_3$ in the TFF: $F_V^2 d_3/(3F_\pi^2)=(-116.2\pm1.8)\cdot10^{-3}$ for this work and $F_V^2 d_3/(3F_\pi^2)=(-110.7\pm8.3)\cdot10^{-3}$ for that in ref.~\cite{us}, which show a good agreement between both determinations.\\

 The study of the so-called off-shell pole contribution to $a_\mu^{HLbL}$ will not be discussed in this article. The $P$--TFF with an off-shell pseudo-Goldstone ($p^2\neq m_P^2$) is by construction an ill-defined quantity (arbitrary off-shell contributions can be obtained through pseudoscalar field redefinitions). One should actually rather analyze the Green's function of four electromagnetic currents $\bra T\{ J^\mu_{\rm EM}(0)  J^\nu_{\rm EM}(x)  J^\alpha_{\rm EM}(y)  J^\beta_{\rm EM}(z)\} \ket $, which is free of these ambiguities. However, its study within the R$\chi$T framework is a more cumbersome problem than just accounting for the pseudo-Goldstone tree-level exchanges provided by the TFFs, and has been postponed for a future work.

 {
 \subsection{Further error analysis}\label{furthererrors}
 
 As said previously, only the leading order terms in $1/N_C$ have been considered in the computation of $a_\mu^{P,HLbL}$. However, 
 one must include a non-zero width in the vector resonance propagators 
 in order to get finite values for the branching fractions of subsection~\ref{subsec:BR}. 
 In particular, the $\rho$ meson width plays the most important role.   
 Intermediate $\pi\pi$ and $K\overline{K}$ loops account for the main NLO corrections in $1/N_C$ to the $\rho$ propagator~\cite{GomezDumm:2000fz}:  
 \begin{equation}
  M_\rho^2 - q^2 \,\, \longrightarrow\,\, 
  M_\rho^2-q^2 +\frac{q^2M_\rho^2}{96\pi^2F_\pi^2}\left(A_\pi(q^2)+\frac{1}{2} A_K(q^2) \right),
  \label{eq:prop-NLO}
 \end{equation}
 where
 \begin{equation}
  A_P(q^2)\, =\,  \ln \frac{m_P^2}{M_\rho^2}+8\frac{m_P^2}{q^2}-\frac{5}{3}+\sigma_P^3(q^2)\ln\left(\frac{\sigma_P(q^2)+1}{\sigma_P(q^2)-1}\right),
 \label{eq:selfenergy-loop}
 \end{equation}
 being $\sigma_P(q^2)=\sqrt{1-\frac{4m_P^2}{q^2}}$. Note that $A_P(q^2)$ is real for $q^2< 4 m_P^2$. Thus, the $\rho$~propagator 
 provided by eq. (\ref{eq:prop-NLO}) is real in the whole space-like region $q^2<0$.

We perform the NLO replacement in the $\rho$ propagators given in eq. (\ref{eq:prop-NLO}) of the functions $g_{V_i}(q^2)$ in app.~\ref{app:Formulae},  
which enter in the $a_\mu^{P,HLbL}$ integral representation~\cite{Knecht}. The $\omega$ and $\phi$ propagators in the $g_{V_i}(q^2)$ 
are left unchanged. Likewise, we keep $f(q^2)=0$, as it is found at LO in $1/N_C$ after imposing the short distance constraints.   
This leads to a decreasing in the theoretical prediction, $a_\mu^{P,HLbL}|_{\rm LO+NLO} - a_\mu^{P,HLbL}|_{\rm LO}= -0.09\cdot10^{-10}$, 
essentially dominated by the contribution in the virtuality range $[0.1,1]$~GeV$^2$, as before.  
This is, nonetheless, just one of the possible NLO corrections in $1/N_C$ to the anomalous magnetic moment. 
One-loop modifications to the $\pi^0 VV'$ vertex can be, e.g., equally important in the space-like domain 
and may lead to a positive contribution to $a_\mu^{P,HLbL}$. Thus, we take the absolute value of this shift as a crude 
estimate of the $1/N_C$ effects: 
\begin{equation}
(\Delta a_\mu^{P,HLbL})_{\rm 1/N_C}\,\, =\,\, \pm 0.09\cdot10^{-10}\, .
\end{equation}

 Other source of unaccounted error originates in the lightest meson dominance assumption of the present work.  
 As shown in \cite{Knecht}, a TFF with only one vector multiplet fails to reproduce at the same time the correct asymptotic behaviour
 $q^2 \mF_{P\gamma^\star\gamma^\star}(0,q^2)= - 2 F$~\cite{BL} and 
 $q^2 \mF_{P\gamma^\star\gamma^\star}(q^2,q^2)= - 2 F/3$~\cite{Nesterenko} for $q^2\to -\infty$.   
 In our one-multiplet study, we were able to reproduce $\mF_{P\gamma^\star\gamma^\star}(0,q^2)\approx  - 2 F/q^2$ at large momentum transfer 
 but the doubly off-shell form-factor behaved like $\mF_{P\gamma^\star\gamma^\star}(q^2,q^2)\stackrel{q^2\to\-\infty}{\sim} \cO(q^{-4})$.  
 However, both QCD limits can be correctly recovered by considering a second multiplet of vector resonances. In the chiral limit,  
 the $\pi^0$--TFF including $\rho$ and $\rho'$ takes then the form  
\begin{eqnarray} 
\mF_{\pi^0\gamma^\star\gamma^\star}(q_1^2,q_2^2) &=&  
\frac{-1}{12\pi^2 F \left(M_{\rho }^2-q_1^2\right)
   \left(M_{\rho }^2-q_2^2\right) \left(M_{\rho '}^2- q_1^2\right) \left(M_{\rho '}^2-q_2^2\right)}
\nn\\
&&
\hspace*{-3cm}\times
\bigg[      -q_1^2 q_2^2 \left(N_C M_{\rho '}^4-48 \pi ^2 F^2 M_{\rho '}^2+4 \pi ^2 F^2 \left(q_1^2+q_2^2\right)\right) 
\nn\\
&&\hspace*{-2.25cm}+N_C M_{\rho }^4 M_{\rho'}^4-8 \pi ^2 F^2 M_{\rho }^2 \left(3 \left(q_1^2+q_2^2\right) M_{\rho '}^2-q_1^2 q_2^2\right)  
\nn\\
&&\hspace*{-1.25cm}+ 64\pi^2 F_\rho^2 d_3^{(\rho,\rho)} M_\rho^2  q_1^2 q_2^2 \left(1- \frac{M_{\rho '}^2}{M_\rho^2} \right)^2
\nn\\
&&\hspace*{-1.25cm}
- \frac{16\pi^2 \sqrt{2} F_\rho c_{125}^{(\rho)}}{M_\rho}  q_1^2 q_2^2 \left(q_1^2-q_2^2\right){}^2 
\left(1-\frac{M_{\rho '}^2}{ M_{\rho }^2  }\right) \bigg]\, .
\label{eq:2R-chiral-TFF}
\end{eqnarray}
 The Lagrangian operators for these second vector multiplet have the same structure as those in Tables \ref{tab:VJP} and \ref{tab:VVP} 
 involving couplings $c_j^{(V)}$ and $d_j^{(V,V')}$, with the indices $V,V'\in\{\rho,\rho'\}$ 
 indicating the resonances in the vertex. In the chiral and large--$N_C$ limits all the vectors in a multiplet are degenerate, 
 so we generically denote a resonance from the first (second) multiplet as $\rho$ ($\rho'$).  
 The comparison of the one-vector-multiplet and two--vector-multiplet TFF provides 
 a rough estimate of the uncertainty introduced in $a_\mu^{\pi^0,HLbL}$ 
 by the incorrect $\mF_{\pi^0\gamma^\star\gamma^\star}(q^2,q^2)$ asymptotic behaviour. 
 Accordingly, we compared the predictions from the chiral limit TFF in eqs.~(\ref{eq:1R-chiral-TFF}) and~(\ref{eq:2R-chiral-TFF}) 
 for  $a_\mu^{\pi^0,HLbL}$ . 
 After imposing that the form-factor follows exactly the QCD asymptotic behaviour from refs.~\cite{BL,Nesterenko},  
 the two-multiplet TFF only depends on two free resonance couplings, which here have been chosen to be $c_{125}^{(\rho)}$ and $d_3^{(\rho,\rho)}$ 
 (the superindex shows the multiplet to which it couples to).   
 One of the short distance conditions is that 
 $F_\rho c_{125}^{(\rho)}/M_\rho^3 + F_{\rho'} c_{125}^{(\rho')}/M_{\rho'}^3    =0$. 
 If one assumes that the lightest multiplet dominates in this high-energy relation, then one would expect that each multiplet contribution 
 cancels on its own, yielding $ c_{125}^{(\rho)}= c_{125}^{(\rho')}=0$, in agreement with the constraint for $c_{125}^{(\rho)}$ 
 in the single-resonance analysis.  
%
%
 Thus, we will use this and the condition $F_\rho^2 d_3^{(\rho,\rho)}= - N_C M_\rho^2/(64\pi^2)$ from eq.~(\ref{eq:new-SD-rels}) in sec.~\ref{SD}.  
 We have taken the chiral limit $\pi^0$--TFF and the inputs $M_\rho=0.77$~GeV, $M_{\rho'}=1.45$~GeV and $F\simeq F_\pi$ 
 in the $a_\mu^{\pi^0,HLbL}$ integral representation~\cite{Knecht}, 
 while keeping the physical pion mass in the integration kernels.  
 Hence, we observe that the second vector multiplet increases $a_\mu^{\pi^0,HLbL}$ by an amount
  $\sim 0.5 \cdot 10^{-10}$.
 This increasing is not unexpected: since our single-resonance $\mF_{\pi^0\gamma^\star\gamma^\star}(q^2,q^2)$ vanishes too fast at high energies,   
 it provides an underrated prediction in comparison with the asymptotic QCD TFF and, consequently, an underrated anomalous 
 magnetic moment prediction. We therefore expect that all modifications of the latter TFF that arrange its asymptotic behaviour will push 
 $a_\mu^{P,HLbL}$ upwards. We thus have the rough estimate,     
 \begin{equation}
  (\Delta a_\mu^{P,HLbL})_{\rm asym} \,\, =\,\, {}^{+0.5}_{-0}\,\cdot10^{-10}\, .
 \end{equation}
%
 We assume the correction to the $\pi^0$ contribution as the dominant effect and neglect here the corrections in $a_\mu^{\eta,HLbL}$ 
 and $a_\mu^{\eta',HLbL}$~\cite{Knecht}.   
 A more rigourous and detailed analysis will be performed in a future work~\cite{NextPaper},
 where we will consider all the short distance constraints beyond the chiral limit 
 and extract the free parameters from a fit to the experimental data.
 This issue has also been discussed in~\cite{Knecht}:
 the difference between the VMD result (reproduced here by~(\ref{eq:1R-chiral-TFF})
 and failing to fulfill the OPE constraint~\cite{Nesterenko}) and the LMD+V rational approximation (with a good short-distance behaviour) 
 was used to estimate an uncertainty $\Delta a_\mu^{\pi^0,LbL}= \pm 1.0 \cdot10^{-10}$. This error 
 was essentially dominated by the uncertainty in one of the LMD+V parameters, namely $h_2$~\cite{Knecht}. 
 }\\

\section{Conclusions}\label{Concl}

We have given a more accurate description of the form-factor by including terms up to order $m_P^2$, for the first time in a chiral invariant Lagrangian approach with the quark mass corrections introduced covariantly.
In addition to chiral symmetry, the high-energy conditions from the TFF (up to $\cO(m_P^2)$) and the VVP Green's funtion (at $\cO(m_P^0)$) were crucial to fix the various unknown Lagrangian parameters.
These short distance constraints were found to be consistent with previous determinations in the chiral and large--$N_C$ limits. We have also been able to fix various $VJP$ couplings for the first time: $c_1=c_2-c_5=c_3=0$.\\

By fitting to the different sets of data for the TFF of the pseudo-Goldstones we were able to confirm that BaBar data for the $\pi^0$-TFF is not compatible with measurements of the $\eta^{(\prime)}$ transition form-factors.
However, in order to stabilize the fit, some prior distribution for the mixing parameters were  provided.
Otherwise, the fit leads to an $\eta-\eta'$ mixing in strong conflict with current phenomenology --though still yielding a fair $a_\mu^{P,HLbL}$ determination--.
After assuming previous phenomenological determinations~\cite{Kaiser:2000gs,eta-etap,mixing} as inputs, we obtain as a by-product of our analysis, the correlated $\eta-\eta'$ mixing parameters, given in table~\ref{tab:mixing}, which may be useful for future analyses involving these mesons.
The contribution from the four mixing inputs $f_{8/0}^{\rm pheno}$ and $\theta_{8/0}^{\rm pheno}$ in eqs.~(\ref{eq:mix-input1})--(\ref{eq:mix-input4})  to the $\chi^2$ turns out to be very small ($\Delta\chi^2\sim 1.5$) and yielded values very similar to previous determinations of such parameters, correlated and with a reduced marginal error. These results depended very mildly on the mixing inputs. \\

Comparing our results with other recent determinations, we find that the TFFs presented in this article turn into rather simple expressions after demanding the correct high energy QCD behaviour while, at the same time, they contain the corresponding $\cO(m_P^2)$ corrections and implement chiral symmetry.
We also have obtained values for the parameters that are compatible with those obtained in ref. \cite{Czyz} despite the lack of heavier copies of the resonances in our approach. This shows that, even though these heavier multiplets are not taken into account, our result should be compatible with the $a_\mu^{P,HLbL}$ considering these resonances.\\

 We have determined the pseudo-Goldstone pole contribution to the $a_\mu^{HLbL}$ with an improved precision,
 \be
a_\mu^{P,HLbL} \,=\, (8.47\, \pm \, 0.16)\, \cdot\, 10^{-10}\, ,
 \ee
with respect to previous works by using a more accurate description of the Resonance Chiral Lagrangian: we have considered corrections  up to $\cO(m_P^2)$ in the form-factor within the lightest $V$ and $P$ resonance multiplet approximation. We have performed some checks which suggest that contributions suppressed by higher powers of $m_P^2$ are negligible at the present level of precision. \\

Future works will be directed towards an improved estimate of the impact from higher resonance multiplets and $1/N_C$ corrections, 
assumed negligible 
{in most of this work.   
Nonetheless, a rough estimate of possible further uncertainties was performed in subsection~\ref{furthererrors}, yielding  
 \begin{equation}
  a_\mu^{P,HLbL} \,\, =\,\,  (\, 8.47\, \pm\, 0.16_{\rm sta}\, \pm \, 0.09_{\rm 1/N_C}\, {}^{+0.5}_{-0}{}_{\rm asym}\, )\, \cdot\, 10^{-10}\, ,
 \end{equation}
 where the first error (sta) comes from the fit, the second one ($1/N_C$) from the estimate of NLO effects in $1/N_C$  
 and the last one (asym) stems from the incorrect $\mF_{P\gamma^\star\gamma^\star}(q^2,q^2)$ asymptotic behaviour and the impact 
 of higher resonance multiplets.   }
Likewise, a more throrough HLbL determination should go beyond the meson-pole contribution, demanding a future study of the $VVVV$ four-point Green's function.

\section*{Acknowledgements}
Work supported by CONACYT Projects No.~FOINS-296-2016 (‘Fronteras de la Ciencia’) and 250628 (‘Ciencia B\'asica’),
by  the  Spanish  MINECO  Project FPA2016-75654-C2-1-P
and by the  Spanish  Consolider-Ingenio  2010  Programme  CPAN  (CSD2007-00042).
A.~G. acknowledges CONACYT for the support 'Estancia Posdoctoral en el Extranjero': "Investigaci\'on apoyada por el CONACYT".
J.J.S.C. would like to thank Z.H.~Guo for discussions on the $\eta$--$\eta'$ mixing. P.R. acknowledges discussions on the short distance constraints with Bastian Kubis, Andreas Nyffeler, Hans Bijnens and Gilberto Colangelo  during the 'Muon g-2 Theory Initiative Hadronic Light-by-Light working group workshop' held at Univ. of Connecticut, 12-14 March 2018.

\appendix

\section{Wess-Zumino-Witten Lagrangian}
\label{app:WZW}

Here we provide the full Wess-Zumino-Witten Lagrangian~\cite{WZW,Bijnens}:
\bear
S[U,\ell,r]_{\mathrm{WZW}}  &=&  -\,\displaystyle \frac{i N_C}{240 \pi^2}
\int d\sigma^{ijklm} \left\langle \Sigma^L_i
\Sigma^L_j \Sigma^L_k \Sigma^L_l \Sigma^L_m \right\rangle
\label{eq:WZW}
\\
&&\mbox{}
 -\,\displaystyle \frac{i N_C}{48 \pi^2} \int d^4 x\,
\varepsilon_{\mu \nu \alpha \beta}\left( W (U,\ell,r)^{\mu \nu
\alpha \beta} - W ({\bf 1},\ell,r)^{\mu \nu \alpha \beta} \right) \, ,
\nn
\eear
\bear
W (U,\ell,r)_{\mu \nu \alpha \beta}  &=&
\left\langle U \ell_{\mu} \ell_{\nu} \ell_{\alpha}U^{\dagger} r_{\beta}
+ \displaystyle \frac{1}{4} U \ell_{\mu} U^{\dagger} r_{\nu} U \ell_\alpha U^{\dagger}
r_{\beta}
\right.
 \label{eq:WZW2} 
 \\
&& \mbox{}
\hspace*{-0.5cm}+ i U \partial_{\mu} \ell_{\nu} \ell_{\alpha} U^{\dagger} r_{\beta}
 +  i \partial_{\mu} r_{\nu} U \ell_{\alpha} U^{\dagger} r_{\beta}
- i \Sigma^L_{\mu} \ell_{\nu} U^{\dagger} r_{\alpha} U \ell_{\beta}
\nonumber\\
&& \mbox{}
\hspace*{-0.5cm}+ \Sigma^L_{\mu} U^{\dagger} \partial_{\nu} r_{\alpha} U \ell_\beta
 -  \Sigma^L_{\mu} \Sigma^L_{\nu} U^{\dagger} r_{\alpha} U \ell_{\beta}
+ \Sigma^L_{\mu} \ell_{\nu} \partial_{\alpha} \ell_{\beta}
+ \Sigma^L_{\mu} \partial_{\nu} \ell_{\alpha} \ell_{\beta}
\nonumber\\
&&  \left.\mbox{}
\hspace*{-0.5cm} - i \Sigma^L_{\mu} \ell_{\nu} \ell_{\alpha} \ell_{\beta}
+ \displaystyle \frac{1}{2} \Sigma^L_{\mu} \ell_{\nu} \Sigma^L_{\alpha} \ell_{\beta}
- i \Sigma^L_{\mu} \Sigma^L_{\nu} \Sigma^L_{\alpha} \ell_{\beta}
\right\rangle
 - \left( L \leftrightarrow R \right) \,,
\nn
\eear
where
\be
\Sigma^L_\mu = U^{\dagger} \partial_\mu U \, , \qquad\qquad
\Sigma^R_\mu = U \partial_\mu U^{\dagger} \, ,
\label{eq:sima_l_r}
\ee
and
$\left( L \leftrightarrow R \right)$ stands for the interchanges
$U \leftrightarrow U^\dagger $, $\ell_\mu \leftrightarrow r_\mu $ and
$\Sigma^L_\mu \leftrightarrow \Sigma^R_\mu $.
The case with an axial vector singlet as well is somewhat more
complicated and is discussed in ref.~\cite{Kaiser:2000ck}.

\section{The $P$ form-factor in the evaluation of $a_\mu^{HLbL}$}\label{app:Formulae}
 For the evaluation of $a_\mu^{HLbL}$ the form-factor is conveniently written as follows \cite{Knecht}

 \begin{equation}
  \mathcal{F}_{P^0\gamma^\star\gamma^\star}(q_1^2,q_2^2)=\frac{F}{3}\left[f(q_1^2)+\sum_{V_i}\frac{1}{M_{V_i}^2-q_2^2}g_{V_i}(q_1^2)\right],\label{TFFamu}
 \end{equation}
 where, for $P^0=\pi^0$ the $f$ and $g$ functions are given by
 \begin{eqnarray}
  f(q^2)&=&\frac{2}{F F_\pi}\left[-\frac{N_C}{8\pi^2}+32m_\pi^2C_7^{W\star} -\sum_{V_i}\frac{2d_3(F_V+8m_\pi^2\lambda_V)^2}{M_{V_i}^2-q^2}\right.\nonumber\\&&\left.
  -\frac{\sqrt{2}(F_V+8m_\pi^2\lambda_V)}{M_V}\left(2c_{1256}+\sum_{V_i}\frac{c_{1235}^\star m_\pi^2-c_{1256}q^2}{M_{V_i}^2-q^2}\right)\right],
 \end{eqnarray}
 \begin{eqnarray}
  g_{V_i}(q^2)&=&\frac{2}{F F_\pi}\left[\sqrt{2}(F_V+8m_\pi^2\lambda_V)\frac{c_{1256}M_{V_i}^2-c_{1235}^\star m_\pi^2}{M_V}\right.\nonumber\\&&\left.
  +2(F_V+8m_\pi^2\lambda_V)^2\frac{d_3(M_{V_i}^2+q^2)+d_{123}^\star m_\pi^2}{M_{V'_i}^2-q^2}\right],
 \end{eqnarray}
 where $M_{V'_i}=M_\omega\delta_{\rho V_i}+M_\rho\delta_{\omega V_i}$, being $\delta$ the Kronecker delta. Notice that since in the $\pi^0$ decays,
 no photon comes from a $\phi$ meson at the large $N_C$ limit (as can be seen in the values of $C^{d}_{1R}$, $C^{m}_{1R}$, $C^{d}_{2R}$
 and $C^{m}_{2R}$ in tables \ref{tab:C1R} and \ref{tab:C2R}) the sum in the $\pi^0$-TFF must only contain the $\rho$ and $\omega$ resonances.\\\hspace*{1ex}

 For the $\eta$-TFF one gets the following $f$ and $g$ functions
 \begin{eqnarray}
  f(q^2)&=&\frac{2}{F^2}\left\{-\frac{(5C_q-\sqrt{2}C_s)N_C}{24\pi^2}+32C_7^{W\star}\, \frac{5C_qm_\pi^2-\sqrt{2}C_s\Delta_{2K\pi}^2}{3}\right.\nonumber\\&&
  +64C_8^W(2C_qm_\pi^2-\sqrt{2}C_s\Delta^2_{2k\pi})-\frac{2(F_V+8m_\pi^2\lambda_V)^2}{3}\sum_{V_i}\frac{h_{V_i}}{M_{V_i}^2-q^2}
  \nonumber\\&&\left.
  -\frac{\sqrt{2}(F_V+8m_\pi^2\lambda_V)}{3M_V}\left[2c_{1256}(5C_q-\sqrt{2}C_s)+\sum_{V_i}\frac{f_{V_i}(q^2)}{M_{V_i}^2-q^2}\right]
  \right\},\nonumber\\
 \end{eqnarray}
 where
 \begin{eqnarray}
  h_\rho&=& 9C_qd_3,\\
  h_\omega&=& C_qd_3,\\
  h_\phi&=& -2\sqrt{2}C_sd_3,\\
  f_\rho(q^2) &=& C_q\left(c_{1235}^\star m_\eta^2-c_{1256}q^2-8c_3^\star \Delta_{\eta\pi}^2\right),\\
  f_\omega(q^2)&=& 9f_\rho(q^2),\\
  f_\phi(q^2)&=& -2\sqrt{2}C_s\left(c_{1235}^\star m_\eta^2-c_{1256}q^2+8c_3^\star \Delta_{2K\pi\eta}^2\right),
 \end{eqnarray}
 and
 \begin{eqnarray}
  g_\rho (q^2)&=& \frac{2}{F^2}\left[\frac{\sqrt{2}(F_V+8m_\pi^2\lambda_V)C_q}{3M_V}\left( c_{1256}M_\rho^2 -c_{1235}^\star m_\eta^2+8c_3^\star \Delta_{\eta\pi}^2\right)\right.\nonumber\\&&\left.
  +6(F_V+8m_\pi^2\lambda_V)^2C_q\frac{d_3(M_\rho^2+q^2)+d_{123}^\star m_\eta^2-8d_2^\star\Delta_{\eta\pi}^2}{M_\rho^2-q^2}\right],\\
  g_\omega (q^2) &=& \frac{2}{F^2}\left[\frac{3\sqrt{2}(F_V+8m_\pi^2\lambda_V)C_q}{M_V}\left(c_{1256}M_\omega^2-c_{1235}^\star m_\eta^2+8c_3^\star \Delta_{\eta\pi}^2\right)\right.\nonumber\\&&\left.
  +\frac{2(F_V+8m_\pi^2\lambda_V)^2C_q}{3}\left(\frac{d_3(M_\omega^2+q^2)+d_{123}^\star m_\eta^2-8d_2^\star\Delta_{\eta\pi}^2}{M_\omega^2-q^2}\right)\right],\\
  g_\phi (q^2) &=& \frac{2}{F^2}\left\{-\frac{4(F_V+8\Delta_{2K\pi}^2\lambda_V)C_s}{3M_V}\left(c_{1256}M_\phi^2-c_{1235}^\star m_\eta^2-8c_3^\star \Delta_{2K\pi\eta}^2\right)\right.\nonumber\\&&\left.
  -\frac{4\sqrt{2}(F_V+8\Delta_{2K\pi}^2\lambda_V)^2C_s}{3}\frac{d_3(M_\phi^2+q^2)+d_{123}^\star m_\eta^2+8d_2^\star\Delta_{2K\pi\eta}^2}{M_\phi^2-q^2}\right\}.\nonumber\\
 \end{eqnarray}
Note that the $f_{\rho,\omega,\phi}$ and $h_{\rho,\omega,\phi}$ in these equations do not refer to the constants employed in ref.~\cite{Czyz} and discussed in Tables~\ref{tab:Czyz} and~\ref{tab:Czyznum}. \\

 Just as explained for the case of the $\eta'-TFF$ below eq. (\ref{etaTFF}), one can obtain the expressions for the $\eta'$ in the form given by eq. (\ref{TFFamu})
 by replacing $C_q\to C_q'$, $C_s\to -C_s'$ and $m_\eta\to m_\eta'$ including $\Delta_{\eta\pi}^2\to\Delta_{\eta'\pi}^2$ and $\Delta_{2K\pi\eta}^2\to\Delta_{2K\pi\eta'}^2$.

 It is worth noticing that, in agreement with refs. \cite{Kampf, us}, by imposing the short distance constraints, one obtains
\bear
f(q^2) \, =\, 0\, ,
\eear
for the three pseudo-Goldstone bosons, which greatly simplifies the numerical computation of $a_\mu^{P,HLbL}$.

\section{Additional information on the fit correlations}
\label{app:correlations}
We collect in this appendix the correlations between fitted parameters in the two main alternative fits discussed in the paper: i) `fit 1', in which all sets of data are analyzed ; ii) `fit 3', where BaBar $\pi^0$ TFF data is excluded from the fits with $M_V$ and $e^V_M$ fixed to the values from~\cite{MassSplitting}. These can be read in Tables \ref{tab:CorrWB} and \ref{tab:CorrMV}, respectively.

 \begin{table}[!ht]
     \centering\begin{tabular}{ccccccccc}\hline
     &$\mathcal{P}_1$&$\mathcal{P}_2$&$M_V$&$e_m^V$&$\theta_8$&$\theta_0$&$f_8$&$f_0$\\\hline\hline
    $\mathcal{P}_1$& 1 & 0.041 & 0.751 & 0.636 & 0.481 & 0.033 & -0.101 & 0.024\\
    $\mathcal{P}_2$& 0.041 & 1 & 0.174 & 0.317 & -0.169 & -0.413 & -0.087 & -0.789\\
    $M_V$& 0.751 & 0.174 & 1 & 0.502 & 0.416 & -0.021 & -0.106 & 0.005\\
    $e_m^V$& 0.636 & 0.317 & 0.502 & 1 & -0.134 & 0.064 & -0.186 & -0.021\\
    $\theta_8$& 0.481 & -0.169 & 0.416 & -0.134 & 1 & -0.031 & -0.443 & 0.024\\
    $\theta_0$& 0.033 & -0.413 & -0.021 & 0.064 & -0.031 & 1 & -0.015 & -0.001\\
    $f_8$& -0.101 & -0.087 & -0.106 & -0.186 & -0.443 & -0.015 & 1 & 0.023\\
    $f_0$ & 0.024 & -0.789 & 0.005 & -0.021 & 0.024 & -0.001 & 0.023 & 1\\\hline\hline
     \end{tabular}\caption{Correlation of fitted parameters including the $\pi^0$ form-factor BaBar data, `fit 1'.}\label{tab:CorrWB}
     \end{table}

\begin{table}[!ht]
     \centering\begin{tabular}{ccccccc}\hline
     &$\mathcal{P}_1$&$\mathcal{P}_2$&$\theta_8$&$\theta_0$&$f_8$&$f_0$\\\hline\hline
    $\mathcal{P}_1$& 1 & 0.005 & 0.778 & -0.012 & 0.009 & 0.078\\
    $\mathcal{P}_2$& 0.005 & 1 & 0.021 & 0.467 & 0.010 & -0.861\\
    $\theta_8$&0.778 & 0.021 & 1 & -0.002 & -0.434 & 0.033\\
    $\theta_0$& -0.012 & 0.467 & -0.002 & 1 & 0.033 & -0.020\\
    $f_8$&0.009 & 0.010 & -0.434 & 0.033 & 1 & 0.002\\
    $f_0$&0.078 & -0.861 & 0.033 & -0.020 & 0.002 & 1\\
\hline\hline
     \end{tabular}\caption{Correlation of fitted parameters with fixed $M_V$ and $e_m^V$ excluding the $\pi^0$ form-factor BaBar data, `fit  3'.}\label{tab:CorrMV}
     \end{table}



\begin{thebibliography}{99}

 \bibitem{PDG}
 C. Patrignani {\it et al.} (Particle Data Group), Chin. Phys. C {\bf 40} (2016) 100001.

 \bibitem{Aoyama:2017uqe}
  T.~Aoyama, T.~Kinoshita and M.~Nio,
  Phys.\ Rev.\ D {\bf 97} (2018) no.3,  036001.

 \bibitem{Bijnens:2007pz}
  J.~Bijnens and J.~Prades,
  Mod.\ Phys.\ Lett.\ A {\bf 22} (2007) 767.

 \bibitem{Miller:2007kk}
  J.~P.~Miller, E.~de Rafael and B.~L.~Roberts,
  Rept.\ Prog.\ Phys.\  {\bf 70} (2007) 795.

 \bibitem{Jegerlehner}
 F. Jegerlehner and A. Nyffeler, Phys. Rep. {\bf 477} (2009) 1.

\bibitem{Lindner:2016bgg}
  M.~Lindner, M.~Platscher and F.~S.~Queiroz,
  Phys.\ Rept.\  {\bf 731} (2018) 1.

\bibitem{Eidelman:2007sb}
  S.~Eidelman and M.~Passera,
  Mod.\ Phys.\ Lett.\ A {\bf 22} (2007) 159.

  \bibitem{Eidelman:2016aih}
  S.~Eidelman, D.~Epifanov, M.~Fael, L.~Mercolli and M.~Passera,
  JHEP {\bf 1603} (2016) 140.

 \bibitem{Brookhaven}
 G.W. Bennett {\it et. al.}, Phys. Rev. Lett. {\bf 92} (2004) 161802.

 \bibitem{Aoyama:2012wk}
  T.~Aoyama, M.~Hayakawa, T.~Kinoshita and M.~Nio,
  Phys.\ Rev.\ Lett.\  {\bf 109} (2012) 111808.

 \bibitem{Gnendiger:2013pva}
  C.~Gnendiger, D.~Stöckinger and H.~Stöckinger-Kim,
  Phys.\ Rev.\ D {\bf 88} (2013) 053005.

 \bibitem{Davier:2010nc}
  M.~Davier, A.~Hoecker, B.~Malaescu and Z.~Zhang,
  Eur.\ Phys.\ J.\ C {\bf 71} (2011) 1515
   Erratum: [Eur.\ Phys.\ J.\ C {\bf 72} (2012) 1874].

 \bibitem{Hagiwara:2011af}
  K.~Hagiwara, R.~Liao, A.~D.~Martin, D.~Nomura and T.~Teubner,
  J.\ Phys.\ G {\bf 38} (2011) 085003.

 \bibitem{Prades:2009tw}
  J.~Prades, E.~de Rafael and A.~Vainshtein,
  Adv.\ Ser.\ Direct.\ High Energy Phys.\  {\bf 20} (2009) 303.

 \bibitem{Kurz:2014wya}
  A.~Kurz, T.~Liu, P.~Marquard and M.~Steinhauser,
  Phys.\ Lett.\ B {\bf 734} (2014) 144.

 \bibitem{Colangelo:2014qya}
  G.~Colangelo, M.~Hoferichter, A.~Nyffeler, M.~Passera and P.~Stoffer,
  Phys.\ Lett.\ B {\bf 735} (2014) 90.

  \bibitem{Davier:2017zfy}
  M.~Davier, A.~Hoecker, B.~Malaescu and Z.~Zhang,
  Eur.\ Phys.\ J.\ C {\bf 77} (2017) no.12,  827.

 \bibitem{Keshavarzi:2018mgv}
  A.~Keshavarzi, D.~Nomura and T.~Teubner,
  arXiv:1802.02995 [hep-ph].

 \bibitem{Actis:2010gg}
  S.~Actis {\it et al.} [Working Group on Radiative Corrections and Monte Carlo Generators for Low Energies],
  Eur.\ Phys.\ J.\ C {\bf 66} (2010) 585.

 \bibitem{FNAL}
 Wesley Gohn, FERMILAB-CONF-17-602-PPD, Muon g-2 collaboration, (2017), E-print: arXiv:1801.00084.

 \bibitem{JPARC}
 Hiromi Inuma {\it et al}, Nucl. Instrum. Meth. A {\bf832} (2016) 51.

 \bibitem{Melnikov:2001uw}
  K.~Melnikov,
  Int.\ J.\ Mod.\ Phys.\ A {\bf 16} (2001) 4591.

 \bibitem{Brodsky:1967sr}
  S.~J.~Brodsky and E.~De Rafael,
  Phys.\ Rev.\  {\bf 168} (1968) 1620.

 \bibitem{Gourdin:1969dm}
  M.~Gourdin and E.~De Rafael,
  Nucl.\ Phys.\ B {\bf 10} (1969) 667.

 \bibitem{latticeHVP}
  T.~Blum,
  Phys.\ Rev.\ Lett.\  {\bf 91} (2003) 052001.
  C.~Aubin and T.~Blum,
  Phys.\ Rev.\ D {\bf 75} (2007) 114502.
  C.~Aubin, T.~Blum, M.~Golterman and S.~Peris,
  Phys.\ Rev.\ D {\bf 86} (2012) 054509.
  X.~Feng, S.~Hashimoto, G.~Hotzel, K.~Jansen, M.~Petschlies and D.~B.~Renner,
  Phys.\ Rev.\ D {\bf 88} (2013) 034505.
  M.~Golterman, K.~Maltman and S.~Peris,
  Phys.\ Rev.\ D {\bf 90} (2014) no.7, 074508.
  G.~Bali and G.~Endrődi,
  Phys.\ Rev.\ D {\bf 92} (2015) no.5,  054506.
  C.~Aubin, T.~Blum, P.~Chau, M.~Golterman, S.~Peris and C.~Tu,
  Phys.\ Rev.\ D {\bf 93} (2016) no.5,  054508.
  T.~Blum {\it et al.},
  Phys.\ Rev.\ Lett.\  {\bf 116} (2016) no.23,  232002.
  T.~Blum {\it et al.} [RBC/UKQCD Collaboration],
  JHEP {\bf 1604} (2016) 063
   Erratum: [JHEP {\bf 1705} (2017) 034].
  J.~Bijnens and J.~Relefors,
  JHEP {\bf 1611} (2016) 086.
  M.~Della Morte {\it et al.},
  JHEP {\bf 1710} (2017) 020.
  D.~Giusti, V.~Lubicz, G.~Martinelli, F.~Sanfilippo and S.~Simula,
  JHEP {\bf 1710} (2017) 157.
  S.~Borsanyi {\it et al.} [Budapest-Marseille-Wuppertal Collaboration],
  arXiv:1711.04980 [hep-lat].

\bibitem{Blum:2016lnc}
  T.~Blum, N.~Christ, M.~Hayakawa, T.~Izubuchi, L.~Jin, C.~Jung and C.~Lehner,
  Phys.\ Rev.\ Lett.\  {\bf 118} (2017) no.2,  022005.

\bibitem{Blum:2017cer}
  T.~Blum, N.~Christ, M.~Hayakawa, T.~Izubuchi, L.~Jin, C.~Jung and C.~Lehner,
  Phys.\ Rev.\ D {\bf 96} (2017) no.3,  034515.

\bibitem{Green:2015sra}
  J.~Green, O.~Gryniuk, G.~von Hippel, H.~B.~Meyer and V.~Pascalutsa,
  Phys.\ Rev.\ Lett.\  {\bf 115} (2015) no.22,  222003.

 \bibitem{Bern}
  G.~Colangelo, M.~Hoferichter, M.~Procura and P.~Stoffer,
  JHEP {\bf 1409} (2014) 091;
  G.~Colangelo, M.~Hoferichter, B.~Kubis, M.~Procura and P.~Stoffer,
  Phys.\ Lett.\ B {\bf 738} (2014) 6;
  G.~Colangelo, M.~Hoferichter, M.~Procura and P.~Stoffer,
  JHEP {\bf 1509} (2015) 074;
  Phys.\ Rev.\ Lett.\  {\bf 118} (2017) no.23,  232001;
  JHEP {\bf 1704} (2017) 161.

 \bibitem{Mainz}
  V.~Pauk and M.~Vanderhaeghen,
  Phys.\ Rev.\ D {\bf 90} (2014) no.11,  113012;
  A.~Nyffeler,
  Phys.\ Rev.\ D {\bf 94} (2016) no.5,  053006;
  I.~Danilkin and M.~Vanderhaeghen,
  Phys.\ Rev.\ D {\bf 95} (2017) no.1,  014019;
  F.~Hagelstein and V.~Pascalutsa,
  Phys.\ Rev.\ Lett.\  {\bf 120} (2018) no.7,  072002.

 \bibitem{Blum:2013xva}
  T.~Blum, A.~Denig, I.~Logashenko, E.~de Rafael, B.~Lee Roberts, T.~Teubner and G.~Venanzoni,
  arXiv:1311.2198 [hep-ph].

 \bibitem{deRafael:1993za}
  E.~de Rafael,
  Phys.\ Lett.\ B {\bf 322} (1994) 239.

\bibitem{Knecht:2001qg}
  M.~Knecht, A.~Nyffeler, M.~Perrottet and E.~de Rafael,
  Phys.\ Rev.\ Lett.\  {\bf 88} (2002) 071802.

\bibitem{Gerardin:2016cqj}
  A.~G\'erardin, H.~B.~Meyer and A.~Nyffeler,
  Phys.\ Rev.\ D {\bf 94} (2016) no.7,  074507.

  \bibitem{DispersionRel}
  C.~Hanhart, A.~Kup\'sc, U.-G.~Meißner, F.~Stollenwerk and A.~Wirzba,
  Eur.\ Phys.\ J.\ C {\bf 73} (2013) no.12,  2668
   Erratum: [Eur.\ Phys.\ J.\ C {\bf 75} (2015) no.6,  242];
  M.~Hoferichter, B.~Kubis, S.~Leupold, F.~Niecknig and S.~P.~Schneider,
  Eur.\ Phys.\ J.\ C {\bf 74} (2014) 3180;
  B.~Kubis and J.~Plenter,
  Eur.\ Phys.\ J.\ C {\bf 75} (2015) no.6,  283;
  C.~W.~Xiao, T.~Dato, C.~Hanhart, B.~Kubis, U.-G.~Mei{\ss}ner and A.~Wirzba,
   arXiv:1509.02194 [hep-ph].
 \bibitem{ChPT}
  S.~Weinberg,
  Physica A {\bf 96} (1979) 327.
  J.~Gasser and H.~Leutwyler,
  Annals Phys.\  {\bf 158} (1984) 142;
  Nucl.\ Phys.\ B {\bf 250} (1985) 465.

\bibitem{RCT}
G. Ecker, J. Gasser, A. Pich, E. De Rafael, Nucl. Phys. B{\bf321} (1989) 311;
G. Ecker, J. Gasser, H. Leutwyler, A. Pich, E. De Rafael, Phys. Lett. B{\bf223} (1989) 425.

\bibitem{Cirigliano:2006hb}
  V.~Cirigliano, G.~Ecker, M.~Eidemuller, R.~Kaiser, A.~Pich and J.~Portol\'es,
  Nucl.\ Phys.\ B {\bf 753} (2006) 139.

  \bibitem{MassSplitting}
  V.~Cirigliano, G.~Ecker, H.~Neufeld and A.~Pich,
  JHEP {\bf 0306} (2003) 012.
  %
  Z.~H.~Guo and J.~J.~Sanz-Cillero,
  Phys.\ Rev.\ D {\bf 79} (2009) 096006.

    \bibitem{VVP}
  P.~D.~Ruiz-Femen\'ia, A.~Pich and J.~Portol\'es,
  JHEP {\bf 0307} (2003) 003.


  \bibitem{Kampf}
   K. Kampf and J. Novotn\'y, Phys. Rev. D{\bf84} (2011) 014036.

  \bibitem{us}
   P. Roig, A. Guevara and G. L\'opez Castro, Phys. Rev. D{\bf89} (2014) 073016.

\bibitem{Knecht}
  M.~Knecht and A.~Nyffeler,
  Phys.\ Rev.\ D {\bf 65} (2002) 073034.

  \bibitem{BaBar}
  B.~Aubert {\it et al.} [BaBar Collaboration],
  Phys.\ Rev.\ D {\bf 80} (2009) 052002.


   \bibitem{BL}
  S. J. Brodsky and G. R. Farrar, Phys. Rev. Lett. {\bf31} (1973) 1153; G. P. Lepage and S. J. Brodsky, Phys. Rev. D {\bf22} (1980)
  2157.

  \bibitem{Czyz}
 H. Czy\.z, P. Kisza and S. Tracz, Phys. Rev. D {\bf 97} (2018) 016006.

  \bibitem{NLODD}
  K. Kampf, J. Novotn\'y and P. S\'anchez-Puertas, 
  Phys.\ Rev.\ D {\bf 97} (2018) no.5,  056010.

  \bibitem{Ametller:1991jv}
  L.~Ametller, J.~Bijnens, A.~Bramon and F.~Cornet,
  Phys.\ Rev.\ D {\bf 45} (1992) 986.

 \bibitem{WZW}
  J.~Wess and B.~Zumino,
  Phys.\ Lett.\  {\bf 37B} (1971) 95;
  E.~Witten,
  Nucl.\ Phys.\ B {\bf 223} (1983) 422.

 \bibitem{SDConst}
  P. Roig and J.J. Sanz Cillero, Phys. Lett. B{\bf733} (2014) 158.

   \bibitem{Bijnens}
 J. Bijnens, L. Girlanda and P. Talavera, Eur. Phys. J. C{\bf 23} (2002) 539.

\bibitem{Kaiser:2000gs}
  R.~Kaiser and H.~Leutwyler,
  Eur.\ Phys.\ J.\ C {\bf 17} (2000) 623.


 \bibitem{eta-etap}
%
  T.~Feldmann, P.~Kroll and B.~Stech,
  Phys.\ Rev.\ D {\bf 58} (1998) 114006;
%
  Phys.\ Lett.\ B {\bf 449} (1999) 339.

  \bibitem{largeNC}
  G.~'t Hooft,
  Nucl.\ Phys.\ B {\bf 72} (1974) 461;
 {\bf 75} (1974) 461.
  E.~Witten,
  Nucl.\ Phys.\ B {\bf 160} (1979) 57.

\bibitem{Bernard:1991zc}
  V.~Bernard, N.~Kaiser and U.~G.~Meissner,
  Nucl.\ Phys.\ B {\bf 364} (1991) 283.

\bibitem{SanzCillero:2004sk}
  J.~J.~Sanz-Cillero,
  Phys.\ Rev.\ D {\bf 70} (2004) 094033.

\bibitem{Guo:2014yva}
  Z.~H.~Guo and J.~J.~Sanz-Cillero,
  Phys.\ Rev.\ D {\bf 89} (2014) no.9,  094024.

    \bibitem{mixing}
  J.~Schechter, A.~Subbaraman and H.~Weigel,
  Phys.\ Rev.\ D {\bf 48} (1993) 339;
  A.~Bramon, R.~Escribano and M.~D.~Scadron,
  Eur.\ Phys.\ J.\ C {\bf 7} (1999) 271;
 T. Feldmann, P. Kroll and D. Stech, Phys. Lett. B {\bf 499}, 339 (1999);
%
  T.~Feldmann,
  Int.\ J.\ Mod.\ Phys.\ A {\bf 15} (2000) 159;
  R.~Escribano and J.~M.~Frere,
  JHEP {\bf 0506} (2005) 029.

  \bibitem{Guo:2015xva}
  X.~K.~Guo, Z.~H.~Guo, J.~A.~Oller and J.~J.~Sanz-Cillero,
  JHEP {\bf 1506} (2015) 175.

  \bibitem{Masjuan:2017tvw}
  P.~Masjuan and P.~S\'anchez-Puertas,
  Phys.\ Rev.\ D {\bf 95} (2017) no.5,  054026.

     \bibitem{Nesterenko}
  V. A. Nesterenko and A. V. Radyushkin, Phys. Lett. B{\bf128} (1983) 439;
  %
  V. A. Novikov, Mikhail A. Shifman, A. I. Vainshtein, M. B. Voloshin, V. I. Zakharov, Nucl. Phys. B{\bf237} (1984) 525.

  \bibitem{Guo:2010dv}
  Z.~H.~Guo and P.~Roig,
  Phys.\ Rev.\ D {\bf 82} (2010) 113016.

\bibitem{Guo:2008sh}
  Z.~H.~Guo,
  Phys.\ Rev.\ D {\bf 78} (2008) 033004.


\bibitem{Dumm:2009kj}
  D.~G.~Dumm, P.~Roig, A.~Pich and J.~Portol\'es,
  Phys.\ Rev.\ D {\bf 81} (2010) 034031.

  \bibitem{Dumm:2012vb}
  D.~G\'omez Dumm and P.~Roig,
  Phys.\ Rev.\ D {\bf 86} (2012) 076009.

  \bibitem{CLEO}
  J.~Gronberg {\it et al.} [CLEO Collaboration],
  Phys.\ Rev.\ D {\bf 57} (1998) 33.

  \bibitem{CELLO}
  H.~J.~Behrend {\it et al.} [CELLO Collaboration],
  Z.\ Phys.\ C {\bf 49} (1991) 401.

  \bibitem{Lep}
  M.~Acciarri {\it et al.} [L3 Collaboration],
  Phys.\ Lett.\ B {\bf 418} (1998) 399.
  \bibitem{BABAR:2011ad}
  P.~del Amo Sanchez {\it et al.} [BaBar Collaboration],
  Phys.\ Rev.\ D {\bf 84} (2011) 052001.

  \bibitem{Belle}
  S. Uehara, {\it et al.} (Belle Collaboration), Phys. Rev. D {\bf 86} (2012) 092007.

  \bibitem{Mikhailov:2009kf}
  S.~V.~Mikhailov and N.~G.~Stefanis,
  Nucl.\ Phys.\ B {\bf 821} (2009) 291.

  \bibitem{Roberts:2010rn}
  H.~L.~L.~Roberts, C.~D.~Roberts, A.~Bashir, L.~X.~Guti\'errez-Guerrero and P.~C.~Tandy,
  Phys.\ Rev.\ C {\bf 82} (2010) 065202.

  \bibitem{Brodsky:2011yv}
  S.~J.~Brodsky, F.~G.~Cao and G.~F.~de Teramond,
  Phys.\ Rev.\ D {\bf 84} (2011) 033001.

\bibitem{Bakulev:2011rp}
  A.~P.~Bakulev, S.~V.~Mikhailov, A.~V.~Pimikov and N.~G.~Stefanis,
  Phys.\ Rev.\ D {\bf 84} (2011) 034014.

\bibitem{Brodsky:2011xx}
  S.~J.~Brodsky, F.~G.~Cao and G.~F.~de Teramond,
  Phys.\ Rev.\ D {\bf 84} (2011) 075012.

  \bibitem{Stefanis:2012yw}
  N.~G.~Stefanis, A.~P.~Bakulev, S.~V.~Mikhailov and A.~V.~Pimikov,
  Phys.\ Rev.\ D {\bf 87} (2013) no.9,  094025.

  \bibitem{Bakulev:2012nh}
  A.~P.~Bakulev, S.~V.~Mikhailov, A.~V.~Pimikov and N.~G.~Stefanis,
  Phys.\ Rev.\ D {\bf 86} (2012) 031501.

\bibitem{Raya:2015gva}
  K.~Raya, L.~Chang, A.~Bashir, J.~J.~Cobos-Mart\'inez, L.~X.~Gutiérrez-Guerrero, C.~D.~Roberts and P.~C.~Tandy,
  Phys.\ Rev.\ D {\bf 93} (2016) no.7,  074017.

\bibitem{Eichmann:2017wil}
  G.~Eichmann, C.~Fischer, E.~Weil and R.~Williams,
  Phys.\ Lett.\ B {\bf 774} (2017) 425.

\bibitem{Agaev:2014wna}
  S.~S.~Agaev, V.~M.~Braun, N.~Offen, F.~A.~Porkert and A.~Schäfer,
  Phys.\ Rev.\ D {\bf 90} (2014) no.7,  074019.

\bibitem{Escribano:2015yup}
  R.~Escribano, S.~Gonzàlez-Solís, P.~Masjuan and P.~Sánchez-Puertas,
  Phys.\ Rev.\ D {\bf 94} (2016) no.5,  054033.

\bibitem{Dai:2013joa}
  L.~Y.~Dai, J.~Portol\'es and O.~Shekhovtsova,
  Phys.\ Rev.\ D {\bf 88} (2013) 056001.

 \bibitem{Chen:2012vw}
  Y.~H.~Chen, Z.~H.~Guo and H.~Q.~Zheng,
  Phys.\ Rev.\ D {\bf 85} (2012) 054018.

\bibitem{Czyz:2012nq}
  H.~Czyz, S.~Ivashyn, A.~Korchin and O.~Shekhovtsova,
  Phys.\ Rev.\ D {\bf 85} (2012) 094010.

\bibitem{Bijnens:1995ii}
  J.~Bijnens and E.~Pallante,
  Mod.\ Phys.\ Lett.\ A {\bf 11} (1996) 1069.

\bibitem{Kampf:2006yf}
  K.~Kampf, J.~Novotny and J.~Trnka,
  Eur.\ Phys.\ J.\ C {\bf 50} (2007) 385.


  \bibitem{GomezDumm:2000fz}
  D.~G\'omez Dumm, A.~Pich and J.~Portol\'es,
  Phys.\ Rev.\ D {\bf 62} (2000) 054014.



  \bibitem{NextPaper}
  A. Guevara, P. Roig and J.~J.~Sanz-Cillero, in preparation.

\bibitem{Hayakawa:1997rq}
  M.~Hayakawa and T.~Kinoshita,
  Phys.\ Rev.\ D {\bf 57} (1998) 465.
   Erratum: [Phys.\ Rev.\ D {\bf 66} (2002) 019902].

\bibitem{Bijnens:2001cq}
  J.~Bijnens, E.~Pallante and J.~Prades,
  Nucl.\ Phys.\ B {\bf 626} (2002) 410.

\bibitem{Erler:2006vu}
  J.~Erler and G.~Toledo S\'anchez,
  Phys.\ Rev.\ Lett.\  {\bf 97} (2006) 161801.

\bibitem{Hong:2009zw}
  D.~K.~Hong and D.~Kim,
  Phys.\ Lett.\ B {\bf 680} (2009) 480.

\bibitem{Cappiello:2010uy}
  L.~Cappiello, O.~Cat\`a and G.~D'Ambrosio,
  Phys.\ Rev.\ D {\bf 83} (2011) 093006.

\bibitem{Goecke:2010if}
  T.~Goecke, C.~S.~Fischer and R.~Williams,
  Phys.\ Rev.\ D {\bf 83} (2011) 094006
   Erratum: [Phys.\ Rev.\ D {\bf 86} (2012) 099901].

  \bibitem{Dorokhov:2011zf}
  A.~E.~Dorokhov, A.~E.~Radzhabov and A.~S.~Zhevlakov,
  Eur.\ Phys.\ J.\ C {\bf 71} (2011) 1702.

  \bibitem{Masjuan:2012wy}
  P.~Masjuan,
  Phys.\ Rev.\ D {\bf 86} (2012) 094021.

 \bibitem{Masjuan:2012qn}
  P.~Masjuan and M.~Vanderhaeghen,
  J.\ Phys.\ G {\bf 42} (2015) no.12,  125004.

\bibitem{Escribano:2013kba}
  R.~Escribano, P.~Masjuan and P.~S\'anchez-Puertas,
  Phys.\ Rev.\ D {\bf 89} (2014) no.3,  034014.

\bibitem{Melnikov:2003xd}
  K.~Melnikov and A.~Vainshtein,
  Phys.\ Rev.\ D {\bf 70} (2004) 113006.

\bibitem{Bijnens:2016hgx}
  J.~Bijnens and J.~Relefors,
  JHEP {\bf 1609} (2016) 113.

\bibitem{Asmussen:2017bup}
  N.~Asmussen, A.~G\'erardin, H.~B.~Meyer and A.~Nyffeler,
  arXiv:1711.02466 [hep-lat].

  \bibitem{Hoferichter:2018dmo}
  M.~Hoferichter, B.~L.~Hoid, B.~Kubis, S.~Leupold and S.~P.~Schneider,
  arXiv:1805.01471 [hep-ph].

\bibitem{Kaiser:2000ck}
  R.~Kaiser,
  Phys.\ Rev.\ D {\bf 63} (2001) 076010;
%
  R.~Kaiser and H.~Leutwyler,
  In {\it Adelaide 1998, Nonperturbative methods in quantum field theory} 15-29
  [hep-ph/9806336].

 \end{thebibliography}
\end{document}